\documentclass[fleqn,10pt]{wlscirep}
\title{Phase-coherent solitonic Josephson heat oscillator}

\author[1*]{Claudio Guarcello}
\author[2]{Paolo Solinas}
\author[1]{Alessandro Braggio}
\author[1]{Francesco Giazotto}
\affil[1]{NEST, Istituto Nanoscienze-CNR and Scuola Normale Superiore, Piazza S. Silvestro 12, I-56127 Pisa, Italy}
\affil[2]{SPIN-CNR, Via Dodecaneso 33, I-16146 Genova, Italy}
\affil[*]{claudio.guarcello@nano.cnr.it}

\begin{abstract}
Since its recent foundation, phase-coherent caloritronics has sparkled continuous interest giving rise to numerous concrete applications. This research field deals with the coherent manipulation of heat currents in mesoscopic superconducting devices by mastering the Josephson phase difference. Here, we introduce a new generation of devices for fast caloritronics able to control local heat power and temperature through manipulation of Josephson vortices, i.e., \emph{solitons}. Although most salient features concerning Josephson vortices in long Josephson junctions were comprehensively hitherto explored, little is known about soliton-sustained coherent thermal transport. We demonstrate that the soliton configuration determines the temperature profile in the junction, so that, in correspondence of each magnetically induced soliton, both the flowing thermal power and the temperature significantly enhance. Finally, we thoroughly discuss a fast solitonic Josephson heat oscillator, whose frequency is in tune with the oscillation frequency of the magnetic drive. Notably, the proposed heat oscillator can effectively find application as a tunable thermal source for nanoscale heat engines and coherent thermal machines.
\end{abstract}
\begin{document}

\flushbottom
\maketitle
%
%
\thispagestyle{empty}

\section*{Introduction}

In recent years, the growing demand of fast electronics promoted the thrive of applications based on Josephson vortices, i.e., solitons~\cite{And10,Lik12,Wus18}. Nonetheless, the effective interplay between thermal transport and soliton dynamics is far from being fully explored. Indeed, the influence on the dynamics of solitons of a homogeneous temperature gradient applied along (namely, from one edge of the junction to the other) a long Josephson junctions (LJJs) was earlier studied, both theoretically and experimentally, in Refs.~\cite{Log94,Gol95,Kra97}. Instead, the issue of the soliton-sustained coherent thermal transport in a LJJ as a temperature gradient is imposed across the system (namely, as the electrodes forming the device reside at different temperatures) was exclusively recently addressed in Refs.~\cite{Gua18,GuaSolBra18}. The latter work reports the first endeavour to combine the physics of solitons and phase-coherent caloritronics~\cite{Mak65,Gia06,MarSol14,ForGia17}. This research field deals with the manipulation of heat currents in mesoscopic superconducting devices~\cite{Gia12,Mar14,Mar15,Sol16,ForBla16,Pao17,ForGia17,For17,Pao18,Tim18} by mastering the phase difference of the superconducting order parameter. 
In this framework, the thermal modulation induced by the external magnetic field was first demonstrated in superconducting quantum-interference devices (SQUID)~\cite{GiaMar12,Gia12} and then in short Josephson junctions (JJs)~\cite{Gia13,Mar14}. Moreover, hysteretic behaviours in temperature-biased Josephson devices were also recently discussed in Refs.~\cite{Gua17,GuaSol18}.

\begin{figure}[hb!!]
\center
\includegraphics[width=0.75\columnwidth]{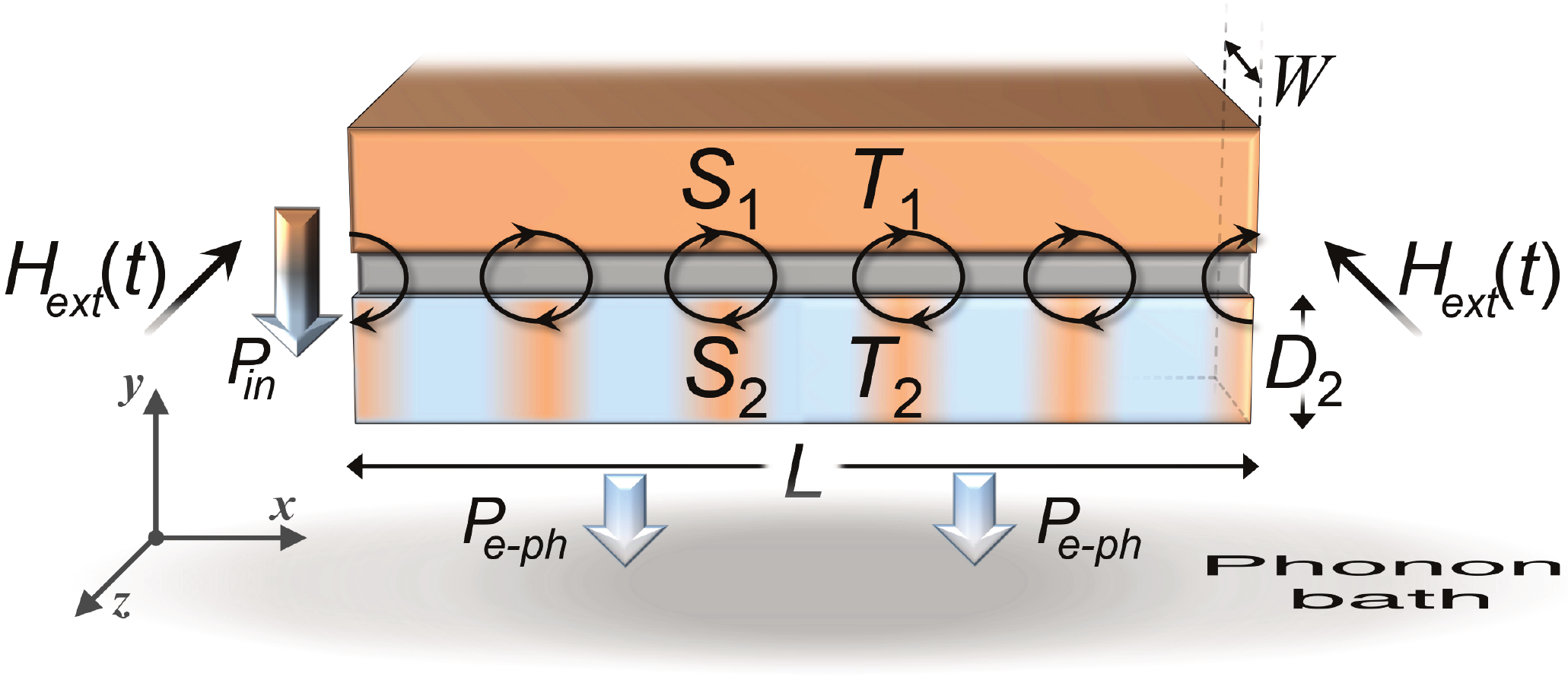}
\caption{\textbf{Fluxon chain in a magnetically driven, thermally biased LJJ.} A \emph{S-I-S} LJJ excited by an external in-plane magnetic field $H_{\text{ext}}(t)$. The length and the width of the junction are $L\gg\lambda_{_{\text{J}}}$ and $W \ll\lambda_{_{\text{J}}}$, respectively, where $\lambda_{_{\text{J}}}$ is the Josephson penetration depth. Moreover, the thickness $D_2 \ll\lambda_{_{\text{J}}}$ of the electrode $S_2$ is indicated. A chain of fluxons, i.e., solitons, along the junction is depicted. The incoming, i.e., $P_{\text{in}}\left ( T_1,T_2,\varphi,V \right )$, and outgoing, i.e., $P_{e\text{-ph}}\left ( T_2,T_{\text{bath}}\right )$, thermal powers in $S_2$ are also represented, for $T_1>T_2(x)>T_{\text{bath}}$.}
\label{Figure01}
\end{figure}

Although little is known about caloritronics effects in these systems, LJJs are still nowadays an active research field, both theoretically~\cite{Gul07,Mon12,Val14,Gua15,Zel15,Pan15,GuaValSpa16,GuaSol17,Hil18,Wus18} and experimentally~\cite{Ooi07,Lik12,Fed12,Mon13,Gra14,Kos14,Fed14,Vet15,Gol17}.
In our work, we explore theoretically the effects of a sinusoidal magnetic drive on the thermal transport across a temperature-biased LJJ (see Fig.~\ref{Figure01}). We show that the behavior of solitons along the system is reflected on the fast evolution of both the heat power through the junction and the temperature of a cold ``thermally floating'' electrode of the device. Accordingly, we observe temperature peaks in correspondence of the magnetically induced solitons. Moreover, in sweeping back and forth the driving field, hysteretic thermal phenomena come to light. Finally, we thoroughly discuss the application of this system as a \emph{heat oscillator}, in which the thermal flux flowing from the junction edge oscillates following the sinusoidal magnetic drive. The dynamical approach that we will address in this work, is essential to establish the performance and the figure of merits of the device, especially when a ``fast'' (with respect to the intrinsic thermalization time scale of the system) magnetic drive is considered.

\section*{Results}
\label{Results}\vskip-0.2cm

{\bf Theoretical modelling. }
We investigate the thermal transport in a temperature-biased long Josephson tunnel junction driven by the external magnetic field, $H_{\text{ext}}(t)$. The electrodynamics of a long and narrow Josephson tunnel junction is usually described by the perturbed sine-Gordon (SG) equation~\cite{Bar82} 
\begin{equation}
\frac{\partial^2 \varphi\big(\widetilde{x},\widetilde{t}\,\big) }{\partial {\widetilde{x}}^2} -\frac{\partial^2 \varphi\big(\widetilde{x},\widetilde{t}\,\big) }{\partial {\widetilde{t}}^{2}}- \sin\Big( \varphi\big ( \widetilde{x},\widetilde{t} \,\big ) \Big )= \alpha\frac{\partial \varphi\big(\widetilde{x},\widetilde{t}\,\big) }{\partial \widetilde{t}}\label{SGeq}
\end{equation}
for the Josephson phase $\varphi$, namely, the phase difference between the wavefunctions describing the carriers in the superconducting electrodes. The time variations of $\varphi$ generates a local voltage drop according to $V(x,t)=\frac{\Phi_0}{2\pi}\frac{\partial \varphi (x,t) }{\partial t}$ (where $\Phi_0= h/2e\simeq2\times10^{-15}\; \textup{Wb}$ is the magnetic flux quantum, with $e$ and $h$ being the electron charge and the Planck constant, respectively).
In previous equations, space and time variables are normalized to the Josephson penetration depth $\lambda_{_{\text{J}}}=\sqrt{\frac{\Phi_0}{2\pi \mu_0}\frac{1}{t_d J_c}}$ and to the inverse of the Josephson plasma frequency $\omega_p=\sqrt{\frac{2\pi}{\Phi_0}\frac{J_c}{C}}$, respectively, i.e., $\widetilde{x}=x/\lambda_{_{\text{J}}}$ and $\widetilde{t}=\omega_pt$. The junction is called ``long'' just because its dimensions in units of $\lambda_{_{\text{J}}}$ are $\widetilde{L}=L/ \lambda_{_{\text{J}}}\gg1$ and $\widetilde{W}=W/ \lambda_{_{\text{J}}}\ll1$ (see Fig.~\ref{Figure01}). 
Here, we introduced the critical current density $J_c$, the effective magnetic thickness $t_d=\lambda_{L,1}\tanh\left ( D_1/2\lambda_{L,1} \right )+\lambda_{L,2}\tanh\left ( D_2/2\lambda_{L,2} \right )+d$~\cite{Gia13,Mar14} (where $\lambda_{L,i}$ and $D_i$ are the London penetration depth and the thickness of the electrode $S_i$, respectively, and $d$ is the insulating layer thickness), and the specific capacitance $C$ of the junction due to the sandwiching of the superconducting electrodes.
The dissipation in the junction is accounted by the damping parameter $\alpha=1/(\omega_pRC)$, with $R$ being the normal-state resistance per area of the junction~\cite{Tin04}. 

The unperturbed SG equation, i.e., $\alpha=0$ in equation~\eqref{SGeq}, admits topologically stable travelling-wave solutions, called \emph{solitons}~\cite{Par93,Ust98}, corresponding to 2$\pi$-twists of the phase, which have the simple analytical expression~\cite{Bar82}
\begin{equation}
\varphi(\widetilde{x}-u\widetilde{t})=4\arctan \left [ \exp \left ( \sigma \frac{\widetilde{x}-\widetilde{x}_0-u\widetilde{t} }{\sqrt{1-u^2}} \right ) \right ],
\label{SGkink}
\end{equation}the Josephson heat interferometer
where $\sigma=\pm1$ is the polarity of the soliton and $u$ is the soliton velocity, measured in units of the Swihart’s velocity $\bar{c}=\lambda_{_{\text{J}}}\omega_p$~\cite{Bar82}.
A soliton has a clear physical meaning in the LJJ framework, since it carries a quantum of magnetic flux, induced by a supercurrent loop surrounding it, with the local magnetic field perpendicularly oriented with respect to the junction length. Thus, solitons in the context of LJJs are usually referred to as fluxons or Josephson vortices. 

The effect on the phase evolution of the driving external magnetic field is accounted by the boundary conditions of equation~\eqref{SGeq},
\begin{equation}
\frac{d\varphi(0,t) }{d\widetilde{x}} = \frac{d\varphi(\widetilde{L},t) }{d\widetilde{x}}=2\frac{H_{\text{ext}}(t)}{H_{c,1}}= H(t).
\label{bcSGeq}
\end{equation}
The coefficient $H_{c,1}=\frac{\Phi_0}{\pi \mu_0 t_d\lambda_{_{\text{J}}}}$ is called the first critical field of a LJJ~\cite{Gold01}, since it is a threshold value above which, namely, for $H_{\text{ext}}(t)>H_{c,1}$, in the absence of bias current solitons penetrate from the junction ends and fill the system with some density depending on both the value of $H(t)$ and the length $L$ of the junction.

The aim of this work is the investigation of the variations of the temperature $T_2$ of the electrode $S_2$ as the magnetic drive is properly swept. Specifically, the modulation of the temperature of the drain ``cold'' electrode is usually obtained by realizing a JJ with a large superconducting electrode, namely, $S_1$, whose temperature $T_1$ is kept fixed, and a smaller electrode, namely, $S_2$, with a small volume and, thereby a small thermal capacity. In this way, the heat transferred significantly affects the temperature $T_2$ of the latter electrode, which is then measured. For the sake of readability, hereafter we will adopt the abbreviated notation in which the $x$ and $t$ dependences are left implicit, namely, $T_2=T_2(x,t)$, $\varphi=\varphi(x,t)$, and $V=V(x,t)$. The electrode $S_2$ can be modelled as a one-dimensional diffusive superconductor at a temperature varying along $L$, so that the evolution of the temperature $T_2$ is given by the time-dependent diffusion equation~\cite{Gua18}
\begin{equation}
\frac{\mathrm{d} }{\mathrm{d} x}\left [\kappa( T_2 ) \frac{\mathrm{d} T_2}{\mathrm{d} x} \right ]+\mathcal{P}_{\text{tot}}\left ( T_1,T_2,\varphi \right )=c_v(T_2)\frac{\mathrm{d} T_2}{\mathrm{d} t},
\label{ThermalBalanceEq}
\end{equation}
where the term
\begin{equation}
\mathcal{P}_{\text{tot}}\left ( T_1,T_2,\varphi \right )=\mathcal{P}_{\text{in}}\left ( T_1,T_2,\varphi,V \right )-\mathcal{P}_{e\text{-ph}}\left ( T_2,T_{\text{bath}}\right )
\label{TotalPower}
\end{equation}
consists of the phase-dependent incoming, i.e., $\mathcal{P}_{\text{in}}\left ( T_1,T_2,\varphi,V \right )$, and the outgoing, i.e., $\mathcal{P}_{e\text{-ph}}\left ( T_2,T_{\text{bath}}\right )$, thermal power densities in $S_2$.
Finally, in equation~\eqref{ThermalBalanceEq}, $c_v(T)=T\frac{\mathrm{d} \mathcal{S}(T)}{\mathrm{d} T}$ is the volume-specific heat capacity, with $\mathcal{S}(T)$ being the electronic entropy density of $S_2$~\cite{Sol16}, and $\kappa(T_2)$ is the electronic heat conductivity~\cite{For17}. We are assuming that the lattice phonons are very well thermalized with the substrate that resides at $T_{bath}$, thanks to the vanishing Kapitza resistance between thin metallic films and the substrate at low temperatures~\cite{Wel94,Gia06}.
The full expressions and the physical meaning of all terms and coefficients in equations~\eqref{ThermalBalanceEq} and~\eqref{TotalPower} are thoroughly discussed in ‘Methods’ section.

To explore the thermal transport in this system, it only remains to include in equation~\eqref{ThermalBalanceEq} the proper phase difference $\varphi(x,t)$ for a LJJ given by numerical solution of equations~\eqref{SGeq} and~\eqref{bcSGeq}, with initial conditions $\varphi(\widetilde{x},0)=d\varphi(\widetilde{x},0)/d\widetilde{t}=0\quad \forall \widetilde{x}\in[0-\widetilde{L}]$.

{\bf Numerical results. }
In the present study, we consider an Nb/AlO$_x$/Nb SIS LJJ characterized by a normal resistance per area $R=50~\Omega~\mu\text{m}^2$ and a specific capacitance $C=50~fF/\mu \text{m}^2$. The linear dimensions of the junction are $L=100\;\mu\text{m}$ for the length, $W=0.5\;\mu\text{m}$ for the width, $D_2=0.1\;\mu\text{m}$ and $d=1\;\text{nm}$ for the thicknesses of $S_2$ and the insulating layer, respectively. 

For the Nb electrode, we assume $\lambda_{L}^0=80\;\text{nm}$, $\sigma_N=6.7\times10^6 \;\Omega^{-1}\text{m}^{-1}$, $\Sigma=3\times10^9\;\textup{W}\textup{m}^{-3}\textup{ K}^{-5}$, $N_F=10^{47}\;\textup{ J}^{-1}\textup{ m}^{-3}$, $\Delta_1(0)=\Delta_2(0)=\Delta=1.764k_BT_c$, with $T_c=9.2\;\text{K}$ being the common critical temperature of the superconductors, and $\gamma_1=\gamma_2=10^{-4}\Delta$.

We impose a thermal gradient across the system, specifically, the bath resides at $T_{\text{bath}}=4.2\;\text{K}$, and $S_1$ is at a temperature $T_1=7\;\text{K}$ kept fixed throughout the computation. This value of the temperature $T_1$ assures the maximal soliton-induced heating in $S_2$, for a bath residing at the liquid helium temperature~\cite{Gua18}. Nonetheless, the soliton-sustained local heating that we are going to discuss could be enhanced by reducing the bath temperature and correspondingly adjusting the temperature $T_1$ of the hot electrode. However, we underline that a lowering of the working temperatures could lead to significantly longer thermal response times~\cite{Gua17}. 

The electronic temperature $T_2(x,t)$ of the electrode $S_2$ is the key quantity to master the thermal transport across the junction, since it floats and can be driven by the external magnetic field. By including the proper temperature-dependence in both the effective magnetic thickness $t_d(T_1,T_2)$ and the Josephson critical current density $J_c( T_1,T_2)$, which varies with the temperatures according to the generalized Ambegaokar and Baratoff formula~\cite{Gia05,Bos16}, we obtain $\lambda_{_{\text{J}}}\simeq7.1\;\mu\text{m}$, $\omega_p\simeq1.3\;\text{THz}$, and $H_{c,1}\simeq5.1\;\text{Oe}$. Moreover, $\alpha\simeq0.3$ corresponding to an underdamped dissipative regime. Anyway, these solitonic parameters weakly depend on the temperature $T_2$, in the range of $T_2$'s values that we will discuss.

\begin{figure*}[!!t]
\center
\includegraphics[width=\textwidth]{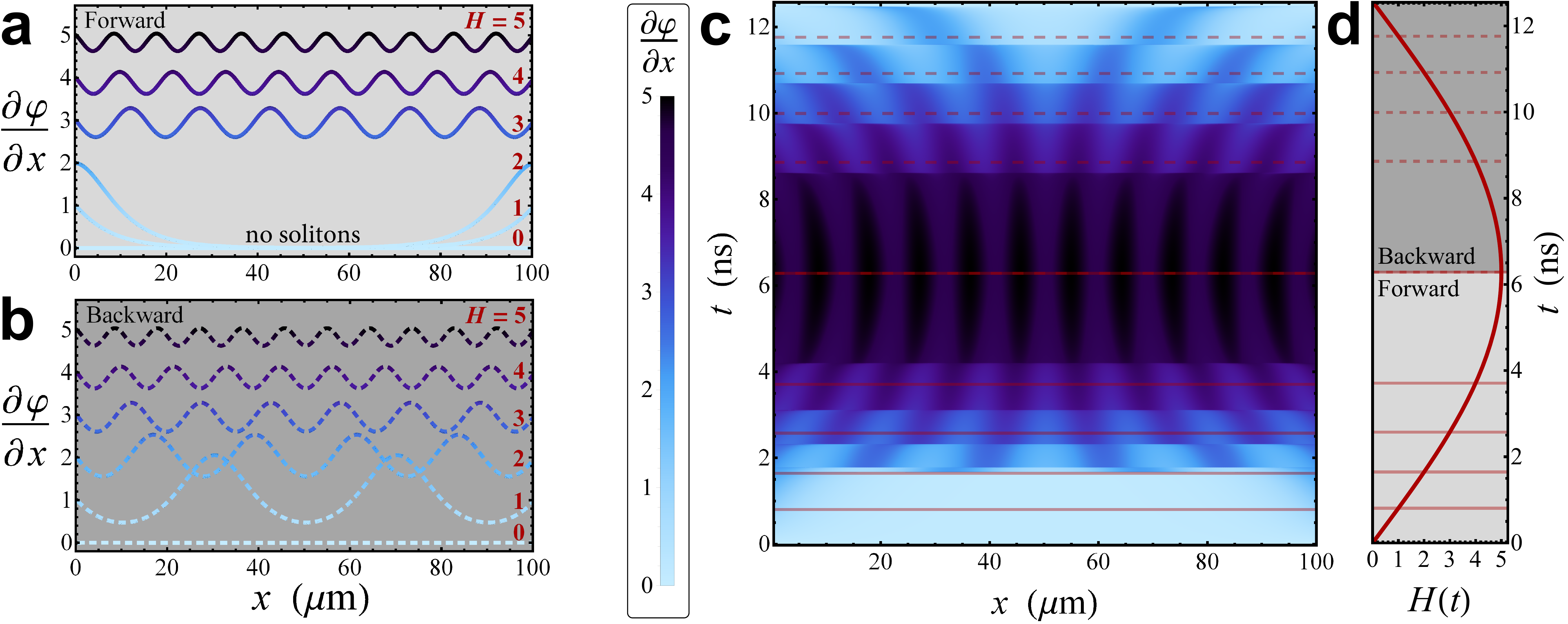}
\caption{\textbf{Soliton configurations as a function of the magnetic drive.} Space derivative of the phase, $\frac{\partial \varphi}{\partial x}$, as a function of $x$ at those times at which the normalized driving field assumes the values $H(t)=\{0,1,2,3,4,5\}$ during the forward [solid lines, panel \textbf{a}] and backward [dashed lines, panels \textbf{b}] sweeps of the drive. In panels \textbf{c} and \textbf{d}, the evolutions of $\frac{\partial \varphi}{\partial x}$ and $H(t)$ are shown, respectively. In the latter panels, the horizontal, red solid and dashed lines indicate the times at which the curves in panels \textbf{a} and \textbf{b} are calculated, respectively.}
\label{Figure02}
\end{figure*}

We use a sinusoidal normalized driving field with frequency $\omega_{\text{dr}}$ and maximum amplitude $H_{\text{max}}$
\begin{equation}\label{drivingfield}
H(t)=H_{\text{max}}\sin(\omega_{\text{dr}} t),
\end{equation}
so that $H(t)$ within a half period is sweeping first forward from 0 to $H_{\text{max}}$ and then backward to 0. In the following, we impose $H_{\text{max}}=5$ and $\omega_{\text{dr}}=0.25\;\text{GHz}$, and we limit ourselves to investigate a single half period of the drive, corresponding to $T_{\text{dr}}/2=4\pi\;\text{ns}$. Anyway, since $\omega_{\text{dr}}\ll\omega_p$, we can image the presented solution for the phase profile as the adiabatic solution of the system. The evolution at multiple periods of the drive can be obtained by simply repeating the presented solution. Interestingly, a driving field sweeping back and forth is expected to give intriguing hysteretic behaviours~\cite{Gua16}.

By increasing the magnetic field, for $H(t)<2$, that means $H_{\text{ext}}(t)<H_{c,1}$ according to equation~\eqref{bcSGeq}, the junction is in the Meissner state, namely, the fluxon-free state of the system~\cite{Cir97,Kup06}. Instead, for a magnetic field above the critical value, i.e., for $H(t)>2$, solitons in the form of fluxons penetrate the LJJ from its ends. However, in this case at a specific value of the magnetic field several solutions, describing distinct configurations with different amount of solitons, may concurrently exist~\cite{Kup06,Kup07,Kup10,Gua16}. The dynamical approach is essential to describe the JJ state when multiple solutions are available. In fact, when the magnetic field increases, at a certain point a configuration with more solitons can be energetically favorable and, thus, the system ``jumps'' from a metastable state to a more stable state. Therefore, the system stays in the present configuration until the following one is energetically more stable. The dynamical approach allows to determine both when the system switches and its new stable state.

The configurations of solitons are well depicted by the space derivative of the phase, $\frac{\partial \varphi(x,t)}{\partial x}$, see Fig.~\ref{Figure02}, since it is proportional to the local magnetic field according to the relation~\cite{Bar82}
\begin{equation}
H_{\text{in}}(x,t)=\frac{\Phi_0}{2\pi\mu_0t_d}\frac{\partial \varphi (x,t) }{\partial x}.
\label{localmagneticfield}
\end{equation}
The spatial distributions of $\frac{\partial \varphi}{\partial x}$ at a few values of $H$, are shown in panels \textbf{a} and \textbf{b} of Fig.~\ref{Figure02}, as the driving field is swept first forward (solid lines) and then backward (dashed lines), respectively. The ripples in these curves indicate fluxons along the junction. For a large applied field the solitons are closely spaced, since the amount of fluxons, e.g., ripples, along the JJ increases by intensifying the magnetic field. In the forward dynamics, for $H\leq2$ the system is in the Meissner state, i.e., no ripples, meaning zero fluxons, and a decaying magnetic field penetrating the junction ends (see Fig.~\ref{Figure02}\textbf{a}). According to the nonlinearity of the problem, for higher fields ($H>2$) the stable solutions are not the trivial superimposition of Meissner and vortex fields, but are rather solitons ``dressed'' by a Meissner field confined in the junction edges~\cite{Kup06}. Hysteresis results looking at the local magnetic field during the backward sweeping of the drive, see Fig.~\ref{Figure02}\textbf{b}. In fact, we observe that forward and backward spatial distributions of $\frac{\partial \varphi(x,t)}{\partial x}$ clearly differ, inasmuch as solitons still persist by reducing the driving field. Nevertheless, for $H(t)=0$ no solitons actually remain within the system in both forward and backward dynamics. 
This hysteretical behavior comes from the multistability of the SG model~\cite{Che94,Kup06,Kup07,Kup10,Gua16,GuaSol17}. In fact, Kuplevakhsky and Glukhov demonstrated that each solution of the SG equation, with a distinct number of solitons, is stable in a broad range of magnetic field values~\cite{Kup06,Kup07,Kup10}. Besides, they observed that these stability regions tend to overlap. Essentially, it means that at a fixed value of the magnetic field different stable solutions, with different amount of solitons along the system, may concurrently exists. Furthermore, it was demonstrated that the longer the junction, the stronger the overlap, and that overlapping decreases by increasing the magnetic field~\cite{Kup06}. This fact not only ensures that the stable solutions cover the whole field range $0\leq H<\infty$, but also proves that hysteresis is an intrinsic property of any LJJ~\cite{Kup06}.
The evolution of a magnetically driven system described by the SG equation can be understood by analyzing the Gibbs free-energy functional and its minimization~\cite{Yug95,Yug99,Kup06,Kup07,Kup10,Gua16}. In fact, by sweeping the magnetic field, the system stays in a current state until the following one is energetically more stable. In this case the system ``jumps'' from a metastable state to a more stable one with a different configuration of solitons.
Interestingly, the hysteresis and the sudden transitions between states with different number of solitons slightly depend also on the damping parameter~\cite{Gua16}. 

The full spatio-temporal evolution of the space derivative of $\varphi$ is displayed in the contour plot in Fig.~\ref{Figure02}\textbf{c}. Alongside, the time evolution of the magnetic drive is shown, see Fig.~\ref{Figure02}\textbf{d}. In both panels, horizontal red solid and dashed lines indicate the times at which the curves in panels \textbf{a} and \textbf{b} are calculated, respectively. In Fig.~\ref{Figure02}\textbf{c}, dark fringes patterns along $x$ indicate solitons along the junction. Furthermore, this figure discloses the transitions between different stable states, when the amount of solitons along the junction changes. As the magnetic field is increased (decreased) the solitons are shifted towards (away from) the center of the junction up to a pair of solitons is symmetrically injected (extracted) from the junction ends. Moreover, we observe that solitons arrange symmetrically and equidistantly along the junction, since the system is centrosymmetric and the solitons with the same polarity tend to repel each other.

%
\begin{figure}[t!!]
\center
\includegraphics[width=0.5\columnwidth]{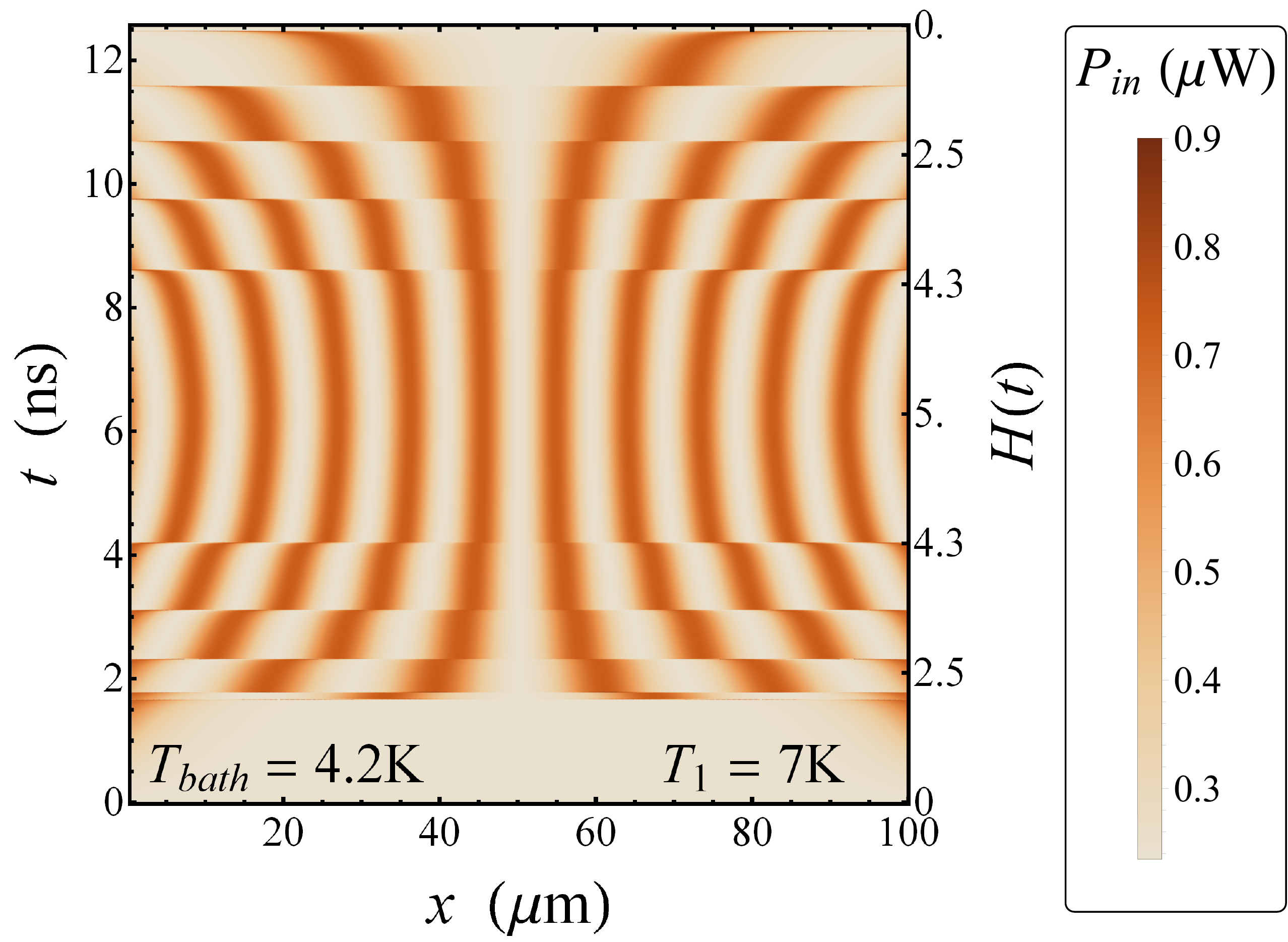}
\caption{\textbf{Phase dependent heat power.} Evolution of the heat power $P_{\text{in}}(x,t)$ at $T_1=7\;\text{K}$ and $T_{\text{bath}}=4.2\;\text{K}$, for $H_{\text{max}}=5$ and $\omega_{\text{dr}}=0.25\;\text{GHz}$. }
\label{Figure03}
\end{figure}
\begin{figure*}[b!!]
\center
\includegraphics[width=\columnwidth]{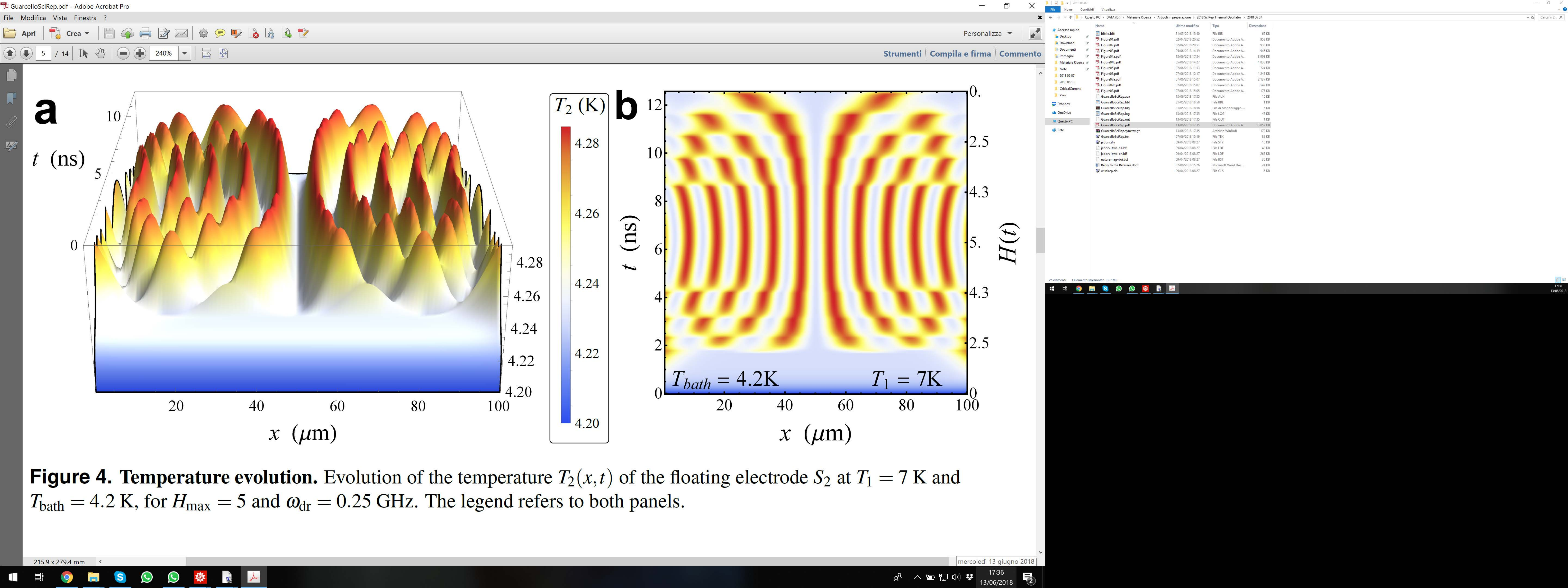}
\caption{\textbf{Temperature evolution.} Evolution of the temperature $T_{2}(x,t)$ of the floating electrode $S_2$ at $T_1=7\;\text{K}$ and $T_{\text{bath}}=4.2\;\text{K}$, for $H_{\text{max}}=5$ and $\omega_{\text{dr}}=0.25\;\text{GHz}$. The legend refers to both panels. }
\label{Figure04}
\end{figure*}

We have seen that the investigation of the full dynamics is crucial to understand the junction behavior, which depends on the full evolution of the system~\cite{Gua16}.
Therefore, it is natural to wonder if also the heat transport throughout the system, and then the temperature of the junction, changes with the history of the system and how it is related to the soliton evolution.

The time and space evolution of the heat power $P_{\text{in}}$ flowing from $S_1$ to $S_2$ is shown in the density plot in Fig.~\ref{Figure03}. In the abscisses of this figure we report the position along the LJJ, whereas on the left we show the time and the corresponding values of the magnetic field are shown on the right.
We observe that solitons locally correspond to clearly enhancements of the heat power $P_{\text{in}}$, namely, the heat current flowing through the junction is significantly supported by a magnetically excited soliton. In fact, the value of the heat power in correspondence of each soliton is $P_{\text{in}}\sim0.9\;\mu\text{W}$, namely, a value three times higher than the power $P_{\text{in}}\sim0.3\;\mu\text{W}$ flowing elsewhere. The configurations of solitons, the sudden transitions between stable states with different amount of solitons, and the hysteretical behavior by sweeping back and forth the driving field are noticeable in Fig.~\ref{Figure03}.

\begin{figure}[t!!]
\center
\includegraphics[width=0.5\columnwidth]{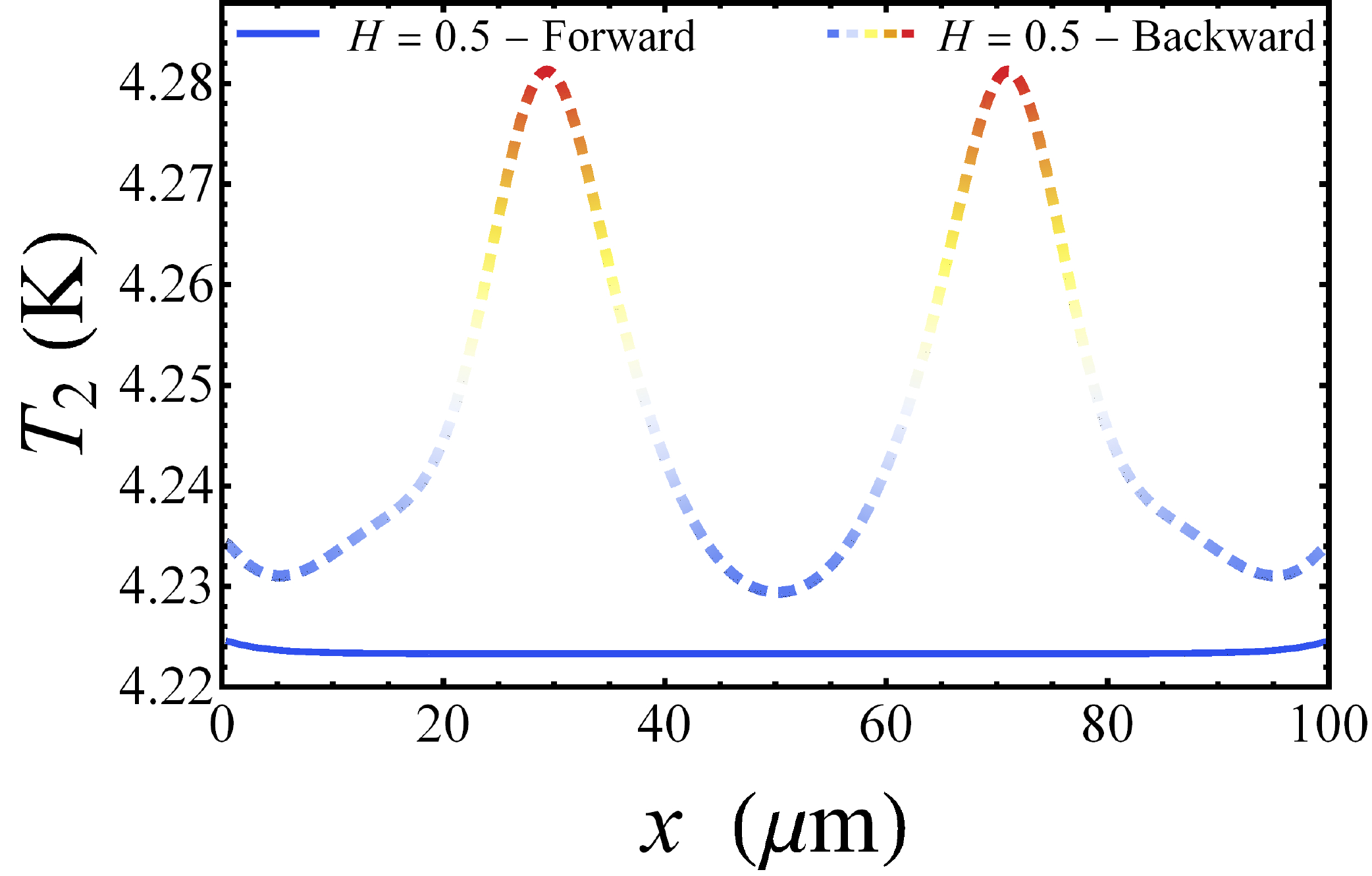}
\caption{\textbf{Thermal hysteresis.} Hysteretic behaviour of the temperature $T_2$ along the junction for $H(t)=0.5$ as the driving field is swept first forward (solid line) and then backward (dashed line).}
\label{Figure05}
\end{figure}
\begin{figure}[b!!]
\center
\includegraphics[width=0.75\columnwidth]{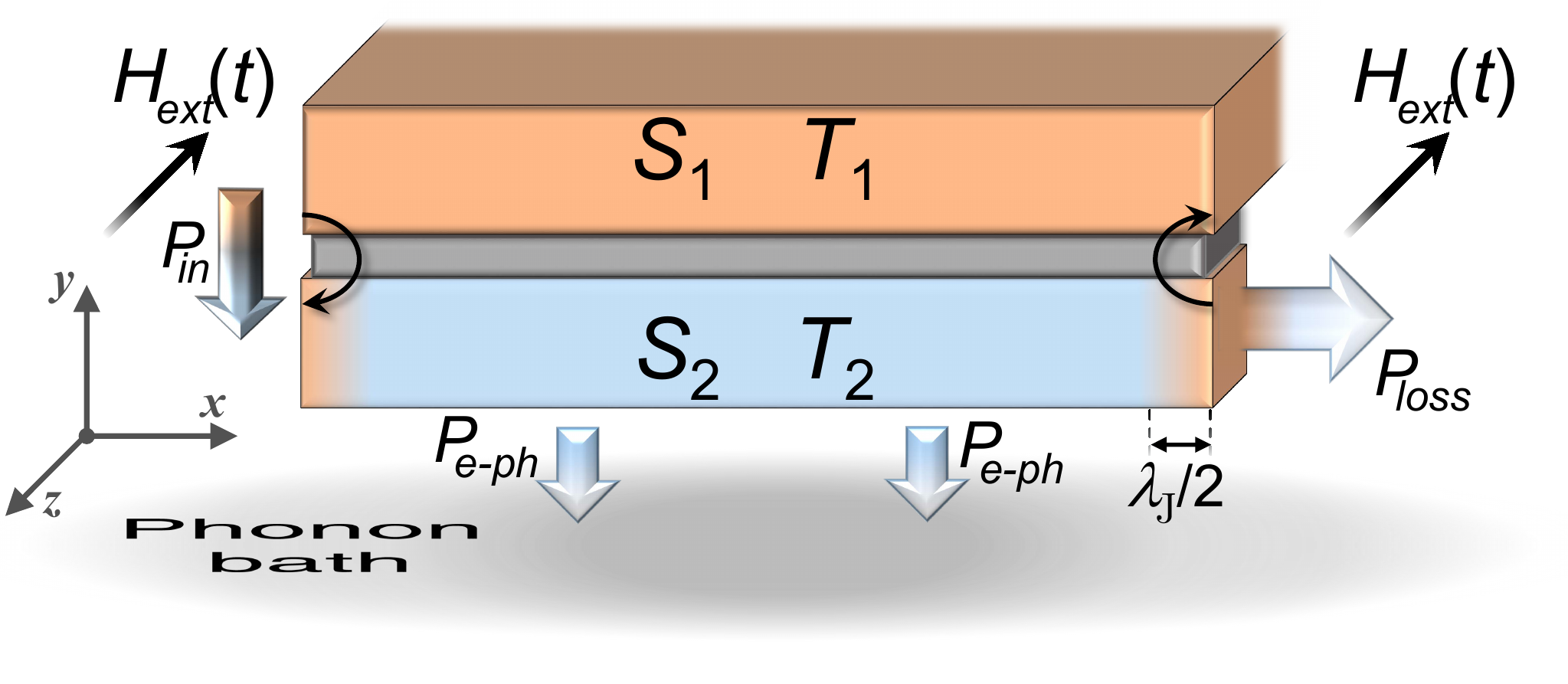}
\caption{\textbf{Magnetically driven long junction operating as a Josephson heat oscillator.} Thermal fluxes in a temperature biased LJJ driven by a sinusoidal magnetic flux, see equation~\ref{drivingfield}, with $H_{\text{max}}=2$. Two solitons confined at the junction edges for $H=H_{\text{max}}$ are also depicted. Besides the incoming and outgoing thermal powers in $S_2$, the thermal power loss $P_{\text{loss}}$ flowing from the right side of $S_2$ is also represented. }
\label{Figure06}
\end{figure}

Finally, the behaviour of the temperature $T_2$ reflects the behavior of the thermal power $P_{\text{in}}$, as it is shown in Fig.~\ref{Figure04}. In panel a of this figure, we observe that when a transition between stable states occurs, the temperature exhibits a locally peaked behavior. We observe that as a soliton set in, it induces a local intense warming-up in $S_2$, so that the temperature of the system locally rapidly approaches the maximum value $T_{2,\text{max}}\simeq4.29\;\text{K}$. Then, when a change in the magnetic field causes a transition to occur, the soliton positions modifies and the temperature adapts to this variation. In fact, the temperature peaks shift according to the new configuration of solitons, see Fig.~\ref{Figure04}\textbf{a}. In this way, for $H=H_{\text{max}}$ several peaks compose the temperature profile, one for each soliton induced by the magnetic field. The contour plot in Fig.~\ref{Figure04}\textbf{b} gives a clear image of the spatio-temporal distribution of $T_2$. In this figure, it is evident how the temperature accurately follows the solitonic dynamics. We note that, for the backward drive, for $H\lesssim1$ two temperature peaks persist, although in the Meissner state (i.e., $H<2$ during the forward sweep of the drive) the whole electrode $S_2$ if roughly thermalized at the same temperature. This thermal hysteresis is evidently highlighted in Fig.~\ref{Figure05}, for $H(t)=0.5$ as the driving field is swept first forward (solid line) and then backward (dashed line).

The results discussed in this work could promptly find application in different contexts. For instance, an alternative method of fluxon imaging in extended JJs could be conceived. Heretofore, the low temperature scanning electron microscopy (LTSEM)~\cite{Hue88,Mal94,Dod97} was confirmed to be an efficient experimental tool for studying fluxon dynamics in Josephson devices. In this technique, a narrow electron beam is used to locally heat a small portion ($\sim\mu\text{m}$) of the junction, in order to locally increase the effective damping parameter. Consequently, the I-V characteristic of the device changes, so that, by gradually moving the electron beam along the junction surface and measuring the voltage, a sort of image of the dynamical state of the LJJ can be created. In our work, we demonstrated that, by imposing a thermal gradient across the junction, the temperature profile of the floating electrode mimics the positions of magnetically-induced solitons. Therefore, our findings can be effectively used for a \emph{thermal} imaging of steady solitons in LJJs through calorimetric measurements~\cite{Gas15,Gia15,Sai16,Zgi17,Wan18}. Moreover, the dynamics discussed so far embodies the thermal router application suggested in Ref.~\cite{Gua18}. 
In fact, we can image a superconducting finger attached in a specific point of $S_2$. Then, by adjusting the external magnetic field, we can induce a specific configuration of solitons along the junction, such to magnetically excite a soliton exactly in correspondence of this finger, with the aim to allow the route of the heat throughout this thermal channel. 
Finally, this device can be used to design a solid-state heat oscillator actively controlled by a magnetic drive. The latter application is carefully discussed in the following section.


%
\begin{figure*}[t]
\center
\includegraphics[width=\textwidth]{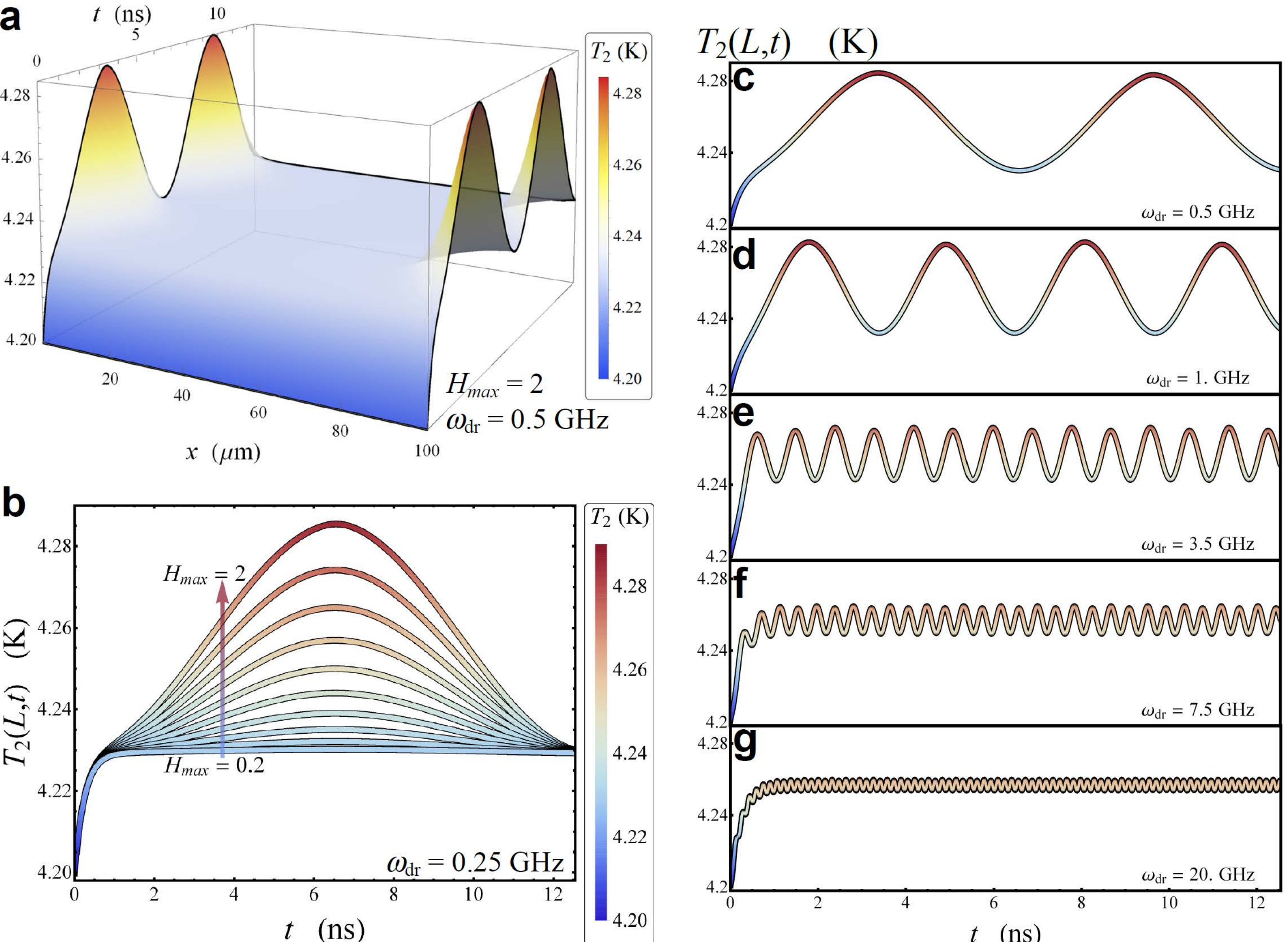}
\caption{\textbf{Thermal oscillations as a function of the magnetic drive.} \textbf{a} Evolution of the temperature $T_{2}(x,t)$ of the floating electrode $S_2$ at $T_1=7\;\text{K}$ and $T_{\text{bath}}=4.2\;\text{K}$, for $H_{\text{max}}=2$, $\omega_{\text{dr}}=0.5\;\text{GHz}$, and $P_{\text{loss}}=0$. \textbf{b} $T_{2}(x,t)$ as a function of $t$ for $x=L$ at a few values of $H_{\text{max}}\in[0.2-2]$ for $\omega_{\text{dr}}=0.25\;\text{GHz}$. \textbf{c}-\textbf{g} $T_{2}(x,t)$ as a function of $t$ for $x=L$, at a few values of $\omega_{\text{dr}}$ for $H_{\text{max}}=2$.}
\label{Figure07}
\end{figure*}

{\bf The Josephson heat oscillator. }
We observe that the sinusoidal magnetic field causes the temperature of both sides of the junction to oscillate, and that, for $H=2$, only two solitons are penetrating the junction, with the center of the solitons being exactly located at the junction edges in $x=\{0,L\}$ (see Fig.~\ref{Figure02}\textbf{a}). Let us now discuss the temperature response through the solitonic dynamics. For $H=2$ we have in $x=\{0,L\}$ the maximum temperature enhancement, since this is the only case in which we can definitively assume a soliton firmly set in a junction end. In fact, the situations for $0<H<2$ can be envisaged by depicting two solitons situated outside the junction, so that by increasing the value of $H$ the solitons moves closer to the junction edges, until their centers are on the borders of junction in $x=\{0,L\}$ for $H = 2$. Instead, for higher fields, i.e., $H>2$, solitons start to penetrate (leave) the junction, so that the temperature of each edge nonlinearly follows the abrupt magnetically driven processes of injection (extraction) of solitons. In light of these remarks, we conceive a heat oscillator based on a temperature biased LJJ driven by a sinusoidal magnetic field with $H_{\text{max}}=2$. 
Specifically, we design to handle the temperature of the right edge of $S_2$, i.e., in $x=L$, with the aim to generate and master a thermal power $P_{\text{loss}}$, which flows from the right side of $S_2$, see Fig.~\ref{Figure06}, which oscillates according to the magnetic drive. 

We first discuss the effects produced by the variations of both the driving amplitude and the frequency on the temperature $T_2(x=L,t)$ when a negligible loss thermal power is assumed, i.e., $P_{\text{loss}}=0$, and then how $P_{\text{loss}}$ affects this temperature.

The modulation of the temperature of $S_2$ due to a sinusoidal drive with $H_{\text{max}}=2$ and $\omega_{\text{dr}}=0.5\;\text{GHz}$, for $P_{\text{loss}}=0$, is displayed in Fig.~\ref{Figure07}\textbf{a}. This figure shows that the enhancement of the temperature is restricted to the junction edges, since for $H\leq2$ there are no solitons inside the system. Moreover, the temperature we are interested in, namely, the temperature of the junction edge in $x=L$, oscillates in tune with the driving field. Specifically, $T_2(L,t)$ shows peaks for $\left | H(t) \right |=H_{\text{max}}$, namely, for $t=T_{\text{dr}}/4$ and $t=3T_{\text{dr}}/4$ (with $T_{\text{dr}}$ being the driving period), and minima for $H(t)=0$. This means that, the thermal oscillation frequency is twice the driving frequency, since the thermal effects are independent on the polarity of the soliton. Accordingly, the frequency requirements of this device are less demanding. This phenomenon allows to discriminate in frequency the magnetic drive from the thermal oscillation. 

Clearly, the oscillatory behavior of the edge temperature $T_2(L,t)$ persists also by reducing $H_{\text{max}}$, see Fig.~\ref{Figure07}\textbf{b} for $\omega_{\text{dr}}=0.25\;\text{GHz}$, although the maximum value of the temperature reduces with decreasing the maximum magnetic drive. The position of the temperature peak is however independent on $H_{\text{max}}$, as it is well demonstrated in Fig.~\ref{Figure07}\textbf{b}. Alternatively, in Figs.\ref{Figure07}\textbf{c}-\textbf{g} the behavior of the temperature $T_2(L,t)$ for $H_{\text{max}}=2$ at a few values of $\omega_{\text{dr}}$ is shown. We note that the temperature oscillation amplitude is drastically damped by increasing the driving frequency, even if the value around which the temperature oscillates is independent on $\omega_{\text{dr}}$. Since the coherent thermal transport is a nonlinear phenomenon, we note that the frequency purity is affected by small corrections inducting vanishingly small spectral components at frequency $\omega_{\text{dr}}$. This effect can be observed as a small beat in the time evolution of $T_2$ (e.g., see Fig.~\ref{Figure07}\textbf{e}).

\begin{figure*}[t!!]
\center
\includegraphics[width=\textwidth]{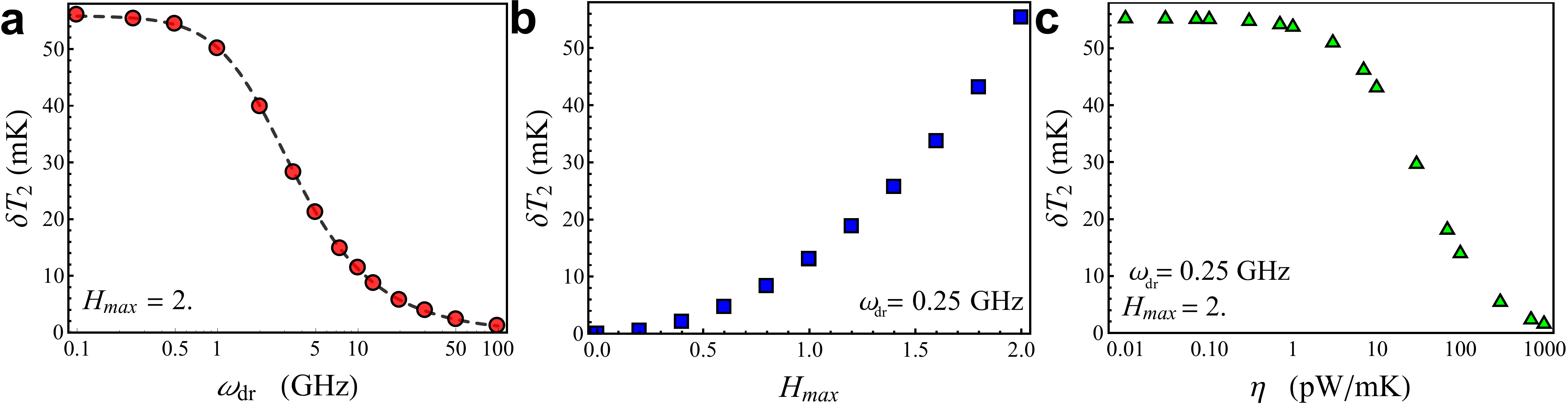}
\caption{\textbf{Figures of merit of the Josephson heat oscillator.} Relevant figures of merit of the heat oscillator, represented by the modulation amplitude, $\delta T_{2}$, as a function of the driving frequency $\omega_{\text{dr}}$ (for $H_{\text{max}}=2$ and $P_{\text{loss}}=0$), the maximum driving amplitude $H_{\text{max}}$ (for $\omega_{\text{dr}}=0.25\;\text{GHz}$ and $P_{\text{loss}}=0$), and the loss thermal conductance $\eta$ (for $H_{\text{max}}=2$ and $\omega_{\text{dr}}=0.25\;\text{GHz}$), see \textbf{a}, \textbf{b}, and \textbf{c}, respectively. In panel \textbf{a} a fitting curve is also shown.}
\label{Figure08}
\end{figure*}

The behavior of the system can be clearly outlined by the $T_2$ modulation amplitude, $\delta T_2$, defined as the difference between the maximum and the minimum values of $T_2(L,t)$ within an oscillation of the drive. In fact, we can define two relevant figures of merit of the thermal oscillator, represented by the modulation amplitude, $\delta T_{2}$, as a function of both the driving frequency $\omega_{\text{dr}}$ (see Fig.~\ref{Figure08}\textbf{a}, for $H_{\text{max}}=2$ and $P_{\text{loss}}=0$) and the maximum driving amplitude $H_{\text{max}}$ (see Fig.~\ref{Figure08}\textbf{b}, for $\omega_{\text{dr}}=0.25\;\text{GHz}$ and $P_{\text{loss}}=0$). 

We look first at the behaviour of $\delta T_2$ by varying the driving frequency $\omega_{\text{dr}}$ (see Fig.~\ref{Figure08}\textbf{a}). We observe that $\delta T_2$ is roughly constant for $\omega_{\text{dr}}\lesssim1\;\text{GHz}$, specifically, $\delta T_2\simeq56\;\text{mK}$. For higher frequencies, the modulation amplitude reduces, going down linearly in the semi-log plot shown in Fig.~\ref{Figure08}\textbf{a}. In fact, as the thermal oscillation frequency becomes comparable to the inverse of the characteristic time scale for the thermal relaxation processes, the temperature is not able to follow the fast driving field. According to Ref.~\cite{GuaSol18}, this thermal response time can be defined as the characteristic time of the exponential evolution by which the temperature locally approaches its stationary value in the presence of a soliton. In Ref.~\cite{Gua18}, for a Nb based junction at the same working temperatures used in this work, a thermal response time roughly equal to $\tau_{\text{th}}\sim0.25\;\text{ns}$ was estimated. 
The modulation amplitude at driving frequency $\omega_{\text{dr}}$ rolls off as $\Big ( 1+(2\omega_{\text{dr}}\tau_{\text{th}})^2 \Big )^{-1/2}$ since $\tau_{\text{th}}$ determines the time scale of the energy exchange between the ensemble and reservoir.
Then, by fitting the $\delta T_2(\omega_{\text{dr}})$ data with the curve $\delta T_{2,0}/\sqrt{ 1+(2\omega_{\text{dr}}\tau_{\text{fit}})^2}$ the parameters $\delta T_{2,0}=\left ( 55.87\pm0.08 \right )\;\text{mK}$ and $\tau_{\text{fit}}=\left ( 0.243\pm0.001 \right )\;\text{ns}$ are estimated (see dashed curve in Fig.~\ref{Figure08}\textbf{a}).
For higher frequencies, other nonlinear effects, related to the finite size of the system, can play a role. Anyway, we are dealing with a regime of tiny temperature modulations at very high frequencies, which is not so significative for practical points of view. 

In Fig.~\ref{Figure08}\textbf{b} the behavior of $\delta T_{2}$ as a function of $H_{\text{max}}$, for $\omega_{\text{dr}}=0.25\;\text{GHz}$, is shown. We observe that, by increasing the maximum drive, the $T_2$ modulation amplitude grows more than linearly. This behavior represents a sort of calibration curve for the thermal oscillator.

Now, we assume a non-vanishing thermal power $P_{\text{loss}}$ flowing throughout the right side of $S_2$, which area is $A=WD_2$. For simplicity, we speculate that this thermal power might depend linearly on the temperature $T_2(L,t)$ according to
\begin{equation}\label{Ploss}
P_{\text{loss}}=\eta \left [ T_2(L,t)-T_{\text{bath}} \right ]=\eta\Delta T_2.
\end{equation}
In this equation, $\eta$ is a thermal conductance, measured in $\text{W}/\text{K}$, that we use as knob to emulate the thermal effectiveness of the load. Thus, another figure of merit of the heat oscillator can be delineated by the $T_2$ modulation amplitude as a function of the coupling constant $\eta$, see Fig.~\ref{Figure08}\textbf{c} for $H_{\text{max}}=2$ and $\omega_{\text{dr}}=0.25\;\text{GHz}$. From this figure we note that $\delta T_2$ is roughly constant for low values of $\eta$, and that the energy outgoing from the right side of $S_2$ affects the temperature of $T_2$ only for $\eta\gtrsim1\;\text{pW}/\text{mK}$. Above this value, the modulation temperature significantly reduces, going to zero for $\eta\sim10^{3}\;\text{pW}/\text{mK}$. 

To estimate the coupling constant threshold value, $\eta_{\text{cr},1}$, above which $\delta T_2$ starts to reduce, we use a scaling argument based on the boundary condition derived by the Fourier law throughout the area $A$ and assuming a temperature drop $\delta T_2$ also along a distance $\lambda_{_{\text{J}}}/2$. Accordingly, we obtain
\begin{equation}\label{criticalcoupliconstant1}
\kappa(T_{2,\text{max}})A\frac{\delta T_2}{\lambda_{_{\text{J}}}/2}=\eta_{\text{cr},1}\Delta T_2,
\end{equation}
from which $\eta_{\text{cr},1}\simeq2.2\;\text{pW/mK}$.

Differently, the value of the critical coupling constant $\eta_{\text{cr},2}$ at witch $\delta T_2\to0$ can be estimated by supposing that the incoming thermal power due to a soliton is entirely balanced by the outgoing power flowing towards both the thermal bath and the right side of $S_2$. We assume that the thermal power induced by a soliton centered in $x=L$ flows through a volume $V_{s}=A\lambda_{_{\text{J}}}/2$. Then, at the equilibrium, from equation\eqref{Ploss}, we obtain
\begin{equation}\label{criticalcoupliconstant2}
\eta_{\text{cr},2}=V_{s}\Bigg ( \frac{\partial \mathcal{P}_{e\text{-ph}}}{\partial T_{2,s}}-\frac{\partial \mathcal{P}_{\text{in}}}{\partial T_{2,s}} \Bigg )=V_{s}\Bigg ( \mathcal{G}(T_{2,s}) -\mathcal{K}(T_{2,s}) \Bigg ),
\end{equation}
where $\mathcal{G}$ and $\mathcal{K}$ are the electron-phonon~\cite{Vir18,GuaSol18} and electron~\cite{Mar14,GuaSol18} thermal conductances, in unit volume, (see ``Methods'' section), and $T_{2,s}$ is the steady temperature in $x=L$. For $T_{2,s}=4.23\;\text{K}$, we obtain $\eta_{\text{cr},2}\simeq1\times10^3\;\text{pW/mK}$.

It is worth noting that the specifications of the proposed thermal oscillator can be tuned by properly choosing the system parameters. For instance, the modulation amplitude could be enhanced by lowering the temperature of the phonon bath and accordingly adjusting the temperature of the hot electrode.
Furthermore, the use of superconductors with higher $T_c$'s gives higher thermalization frequencies, and then it permits to push forward the frequency threshold below which no attenuations of $\delta T_2$ occur. 

The proposed heat oscillator could find application as a temperature controller for heat engines~\cite{Kim11,Ros14,Ros16,Cam16,Mar16}. In fact, in mesoscale and nanoscale systems the precise control of the temperature in a fast time scale is regarded as a difficulty to cope with. Thus, through this system we could be able to definitively master the temperature, which oscillates in a controlled way by a fast magnetic drive. Accordingly, we can envision to build nanoscale heat motors or thermal cycles based on this Josephson heat oscillator.


In summary, in this paper we have thoroughly investigated the effects produced by a time-dependent driving magnetic field on the temperature profile of a long Josephson junction, as a thermal gradient across the system is imposed. A proper magnetic drive induces Josephson vortices, i.e., solitons, along the junction. We showed that the soliton configuration is reflected first on the distribution of heat power flowing through the system and then on the temperature of a cold electrode of the device. In fact, we demonstrated a multipeaked temperature profile, due to the local warming-up of the junction in correspondence of each magnetically excited soliton. Moreover, the study of the full evolution of the system disclosed a clear thermal hysteretic effect as a function of the magnetic drive. We explored a realistic Nb-based setup, where the temperature of the ``hot'' electrode is kept fixed and the thermal contact with a phonon bath at $T_{\text{bath}}=4.2\;\text{K}$ is taken into account. Nevertheless, the soliton-induced heating that we observed can be increased by reducing the temperature of the phonon bath and manipulate by properly control the magnetic drive. 

Finally, we discussed the implementation of a heat oscillator based on this system. 
In the Meissner state, $H<H_{c,1}$, the magnetic drive affects significantly the phase at the junction edges.
In these locations, a clear temperature enhancement is observed. 
Thus, a sinusoidal external magnetic field, with maximum value equal to $H_{c,1}$, causes the edges temperature to oscillate with a frequency twice than the driving field one. 
This phenomenon can be used to conceive a low temperature, field-controlled heat oscillator device based on the thermal diffusion in a Josephson junction, for creating an oscillating heat flux from a spatial thermal gradient between the warm electrode and a cold reservoir, i.e. the phonon bath. The thermal oscillator may have numerous applications, inasmuch as the creation and utilization of an alternating heat flux is applicable to technical systems operating in response to periodic temperature variations, like heat engines, energy-harvesting devices, sensing devices, switching devices, or clocking devices for caloritronics circuits and thermal logic.
Additionally, through proper figures of merit, we discussed the behavior of this heat oscillator by varying both the frequency and the amplitude of the driving field, and also by assuming a non-vanishing loss power flowing towards a thermal load. 
Especially in this context, the dynamical, i.e., fully time-dependent, approach that we used is crucial to understand how the system thermally responds to a fast magnetic drive. For instance, we observed that when the driving frequency becomes comparable to the inverse of the thermalization characteristic times~\cite{GuaSol18}, the system is no longer able to efficiently follow the drive, and the modulation range of the temperature accordingly reduces.

\section*{Methods}
\label{AppA}

\textbf{Thermal Powers. }
In the adiabatic regime, the contributes, in unit volume, to the energy transport in a temperature-biased JJ read~\cite{Gol13}
\begin{equation}\label{Pqp}
\mathcal{P}_{\text{qp}}(T_1,T_2,V)=\frac{1}{e^2RD_2}\int_{-\infty}^{\infty} d\varepsilon \mathcal{N}_1 ( \varepsilon-eV ,T_1 )\mathcal{N}_2 ( \varepsilon ,T_2 )(\varepsilon-eV) [ f ( \varepsilon-eV ,T_1 ) -f ( \varepsilon ,T_2 ) ],
\end{equation}
\begin{equation}
\label{Pcos}
\mathcal{P}_{\cos}( T_1,T_2,V )=\frac{1}{e^2RD_2}\int_{-\infty}^{\infty} d\varepsilon \mathcal{N}_1 ( \varepsilon-eV ,T_1 )\mathcal{N}_2( \varepsilon ,T_2 )\frac{\Delta_1(T_1)\Delta_2(T_2)}{\varepsilon}[ f ( \varepsilon-eV ,T_1 ) -f ( \varepsilon ,T_2 ) ],
\end{equation}
\begin{equation}\label{Psin}
\mathcal{P}_{\sin}(T_1,T_2,V)=\frac{eV}{2\pi e^2RD_2}\iint_{-\infty}^{\infty} d\epsilon_1d\epsilon_2 \frac{\Delta_1(T_1)\Delta_2(T_2)}{E_2}\times\left [\frac{1-f(E_1,T_1)-f(E_2,T_2)}{\left ( E_1+E_2 \right )^2-e^2V^2}\text{+}\frac{f(E_1,T_1)-f(E_2,T_2)}{\left ( E_1-E_2 \right )^2-e^2V^2}\right ]
\end{equation}
where $E_j=\sqrt{\epsilon_j^2+\Delta_j(T_j)^2}$, $f ( E ,T )=1/\left (1+e^{E/k_BT} \right )$ is the Fermi distribution function, $\mathcal{N}_j\left ( \varepsilon ,T \right )=\left | \text{Re}\left [ \frac{ \varepsilon +i\gamma_j}{\sqrt{(\varepsilon +i\gamma_j) ^2-\Delta _j\left ( T \right )^2}} \right ] \right |$ is the reduced superconducting density of state, with $\Delta_j\left ( T_j \right )$ and $\gamma_j$ being the BCS energy gap and the Dynes broadening parameter~\cite{Dyn78} of the $j$-th electrode, respectively.

These equations derives from processes involving both Cooper pairs and quasiparticles in tunneling through a JJ predicted by Maki and Griffin~\cite{Mak65}. In fact, $\mathcal{P}_{\text{qp}}$ is the heat power density carried by quasiparticles, namely, it is an incoherent flow of energy through the junction from the hot to the cold electrode~\cite{Mak65,Gia06}. Instead, the ``anomalous'' terms $\mathcal{P}_{\sin}$ and $\mathcal{P}_{\cos}$ determine the phase-dependent part of the heat transport originating from the energy-carrying tunneling processes involving Cooper pairs and recombination/destruction of Cooper pairs on both sides of the junction.

We note that $\mathcal{P}_{\sin}$, in the temperature regimes we are taking into account, is vanishingly small with respect to both $\mathcal{P}_{\text{qp}}$ and $\mathcal{P}_{\cos}$ contributions, and it can be, in principle, neglected. Moreover, since this term depends on the time derivative of the phase, it could be effective only when the phase rapidly changes, namely, when the soliton enter, or escape, the junction. However, the timescale of the soliton evolution is definitively shorter than the timescales of the driving processes. Consequently, the soliton phase profile follows adiabatically the driving induced by the magnetic field. In this condition, if the number of trapped solitons along the junction is fixed the time scale evolution of the phase is given by the driving process. Anyway, we stress that equation~\eqref{Psin} is a purely reactive contributions~\cite{Gol13,Vir17}, so that in the thermal balance equation~\eqref{ThermalBalanceEq} we have to neglect it. Therefore, the total thermal power density to include in Eq.~\eqref{TotalPower} reads
\begin{equation}\label{Pt}
\mathcal{P}_{\text{in}}( T_1,T_2,\varphi,V)=\mathcal{P}_{\text{qp}}( T_1,T_2,V)-\cos\varphi \;\mathcal{P}_{\cos}( T_1,T_2,V)
\end{equation}

The latter term of the rhs of equation~\eqref{TotalPower}, i.e., $\mathcal{P}_{e\text{-ph}}$, represents the energy exchange, in unit volume, between electrons and phonons in the superconductor and reads~\cite{Pek09}
\begin{eqnarray}\label{Qe-ph}
\mathcal{P}_{e\text{-ph}}&=&\frac{-\Sigma}{96\zeta(5)k_B^5}\int_{-\infty }^{\infty}dEE\int_{-\infty }^{\infty}d\varepsilon \varepsilon^2\textup{sign}(\varepsilon)M_{_{E,E+\varepsilon}}\\\nonumber
&\times& \Bigg\{ \coth\left ( \frac{\varepsilon }{2k_BT_{\text{bath}}}\right ) \Big [ \mathcal{F}(E,T_2)-\mathcal{F}(E+\varepsilon,T_2) \Big ]-\mathcal{F}(E,T_2)\mathcal{F}(E+\varepsilon,T_2)+1 \Bigg\},
\end{eqnarray}
where $\mathcal{F}\left ( \varepsilon ,T_2 \right )=\tanh\left ( \varepsilon/2 k_B T_2 \right )$, $M_{E,{E}'}=\mathcal{N}_i(E,T_2)\mathcal{N}_i({E}',T_2)\left [ 1-\Delta ^2(T_2)/(E{E}') \right ]$, $\Sigma$ is the electron-phonon coupling constant, and $\zeta$ is the Riemann zeta function. 

Finally, in equation~\eqref{ThermalBalanceEq}, $c_v(T)=T\frac{\mathrm{d} \mathcal{S}(T)}{\mathrm{d} T}$ is the volume-specific heat capacity, with $\mathcal{S}(T)$ being the electronic entropy density of $S_2$~\cite{Sol16}
\begin{eqnarray}
\mathcal{S}(T)=-4k_BN_F\int_{0}^{\infty}d\varepsilon \mathcal{N}_2(\varepsilon,T)\left\{ \left [ 1-f(\varepsilon,T) \right ] \ln\left [ 1-f(\varepsilon,T) \right ]+f(\varepsilon,T) \ln f(\varepsilon,T)\right \},\qquad
\label{Entropy}
\end{eqnarray}
and $\kappa(T_2)$ is the electronic heat conductivity~\cite{For17}
\begin{equation}\label{electronicheatconductivity}
\kappa(T_2)=\frac{\sigma_N}{2e^2k_BT_2^2}\int_{-\infty}^{\infty}\mathrm{d}\varepsilon\varepsilon^2\frac{\cos^2\left \{ \text{Im} \left [\text{arctanh} \left (\frac{\Delta(T_2)}{\varepsilon+i\gamma_2} \right )\right ] \right \}}{\cosh ^2 \left (\frac{\varepsilon}{2k_BT_2} \right )},
\end{equation}
with $\sigma_N$ and $N_F$ being the electrical conductivity in the normal state and the density of states at the Fermi energy, respectively.

The first derivative of the heat power densities in equation~\eqref{TotalPower}, calculated at a steady electronic temperature $T_e$, gives the electron-phonon thermal conductance~\cite{Vir18}, in unit volume,
\begin{eqnarray}
\mathcal{G}(T_e)=\frac{\partial \mathcal{P}_{e\text{-ph}}}{\partial T_e} =\frac{5\Sigma}{960\zeta (5)k_B^6T_e^6}\displaystyle\iint_{-\infty}^{\infty}\frac{dEd\varepsilon E\left | \varepsilon \right |^3M^i_{E,E-\varepsilon }}{\sinh\frac{\varepsilon }{2k_BT_e}\cosh\frac{E }{2k_BT_e}\cosh\frac{E-\varepsilon }{2k_BT_e}}
\label{ephconductance}
\end{eqnarray}
and the electron thermal conductance~\cite{Mar14}, in unit volume,
\begin{eqnarray}
\mathcal{K}(T_e)=\frac{\partial \mathcal{P}_{\text{in}}}{\partial T_e}=\frac{1}{2e^2k_BT_e^2RD_2}\displaystyle\int_{0}^{\infty}\frac{d\varepsilon \varepsilon^2}{\cosh^2\frac{\varepsilon }{2k_BT_e}}\Big [\mathcal{N}_1(\varepsilon,T_e)\mathcal{N}_2(\varepsilon,T_e)-\mathcal{M}_1(\varepsilon,T_e)\mathcal{M}_2(\varepsilon,T_e)\cos\varphi\Big ],\qquad
\label{econductance}
\end{eqnarray}
where $\mathcal{M}_j\left ( \varepsilon ,T \right )=\left |\text{Im}\left [\frac{ - i\Delta _j\left ( T \right )}{\sqrt{\left (\varepsilon+ i\gamma_j \right ) ^2-\Delta _j\left ( T \right )^2}} \right ] \right |$.

To estimate $\eta_{cr,2}$ through equation~\eqref{criticalcoupliconstant2}, we assume $\varphi=\pi$ in equation\eqref{econductance}, since the center of the soliton, in correspondence of which $\varphi=\pi$, is placed exactly in the junction edge $x=L$. 


\begin{thebibliography}{10}
\expandafter\ifx\csname url\endcsname\relax
 \def\url#1{\texttt{#1}}\fi
\expandafter\ifx\csname urlprefix\endcsname\relax\def\urlprefix{URL }\fi
\expandafter\ifx\csname doiprefix\endcsname\relax\def\doiprefix{DOI }\fi
\providecommand{\bibinfo}[2]{#2}
\providecommand{\eprint}[2][]{\url{#2}}

\bibitem{And10}
\bibinfo{author}{Anders, S.} \emph{et~al.}
\newblock \bibinfo{journal}{\bibinfo{title}{European roadmap on superconductive
 electronics--status and perspectives}}.
\newblock {\emph{\JournalTitle{Physica C: Superconductivity}}}
 \textbf{\bibinfo{volume}{470}}, \bibinfo{pages}{2079 -- 2126}
 (\bibinfo{year}{2010}).

\bibitem{Lik12}
\bibinfo{author}{Likharev, K.~K.}
\newblock \bibinfo{journal}{\bibinfo{title}{Superconductor digital
 electronics}}.
\newblock {\emph{\JournalTitle{Physica (Amsterdam)}}}
 \textbf{\bibinfo{volume}{482C}}, \bibinfo{pages}{6 -- 18}
 (\bibinfo{year}{2012}).

\bibitem{Wus18}
\bibinfo{author}{Wustmann, W.} \& \bibinfo{author}{Osborn, K.~D.}
\newblock \bibinfo{journal}{\bibinfo{title}{Reversible fluxon logic:
 Topological particles allow gates beyond the standard adiabatic limit}}.
\newblock {\emph{\JournalTitle{arXiv preprint arXiv:1711.04339}}}
 (\bibinfo{year}{2018}).

\bibitem{Log94}
\bibinfo{author}{Logvenov, G.}, \bibinfo{author}{Vernik, I.},
 \bibinfo{author}{Goncharov, M.}, \bibinfo{author}{Kohlstedt, H.} \&
 \bibinfo{author}{Ustinov, A.}
\newblock \bibinfo{journal}{\bibinfo{title}{Dynamics of Josephson vortices in a
 temperature gradient}}.
\newblock {\emph{\JournalTitle{Phys. Lett. A}}} \textbf{\bibinfo{volume}{196}},
 \bibinfo{pages}{76 -- 82} (\bibinfo{year}{1994}).

\bibitem{Gol95}
\bibinfo{author}{Golubov, A.~A.} \& \bibinfo{author}{Logvenov, G.~Y.}
\newblock \bibinfo{journal}{\bibinfo{title}{Motion of a Josephson vortex under
 a temperature gradient}}.
\newblock {\emph{\JournalTitle{Phys. Rev. B}}} \textbf{\bibinfo{volume}{51}},
 \bibinfo{pages}{3696--3700} (\bibinfo{year}{1995}).

\bibitem{Kra97}
\bibinfo{author}{Krasnov, V.~M.}, \bibinfo{author}{Oboznov, V.~A.} \&
 \bibinfo{author}{Pedersen, N.~F.}
\newblock \bibinfo{journal}{\bibinfo{title}{Fluxon dynamics in long Josephson
 junctions in the presence of a temperature gradient or spatial
 nonuniformity}}.
\newblock {\emph{\JournalTitle{Phys. Rev. B}}} \textbf{\bibinfo{volume}{55}},
 \bibinfo{pages}{14486--14498} (\bibinfo{year}{1997}).

\bibitem{Gua18}
\bibinfo{author}{Guarcello, C.}, \bibinfo{author}{Solinas, P.},
 \bibinfo{author}{Braggio, A.} \& \bibinfo{author}{Giazotto, F.}
\newblock \bibinfo{journal}{\bibinfo{title}{Solitonic Josephson thermal
 transport}}.
\newblock {\emph{\JournalTitle{Phys. Rev. Applied}}}
 \textbf{\bibinfo{volume}{9}}, \bibinfo{pages}{034014} (\bibinfo{year}{2018}).


\bibitem{GuaSolBra18}
\bibinfo{author}{Guarcello, C.}, \bibinfo{author}{Solinas, P.},
 \bibinfo{author}{Braggio, A.} \& \bibinfo{author}{Giazotto, F.}
\newblock \bibinfo{journal}{\bibinfo{title}{Solitonic thermal transport in a
 current biased long josephson junction}}.
\newblock {\emph{\JournalTitle{arXiv preprint arXiv:1805.05685}}}
 (\bibinfo{year}{2018}).


\bibitem{Mak65}
\bibinfo{author}{Maki, K.} \& \bibinfo{author}{Griffin, A.}
\newblock \bibinfo{journal}{\bibinfo{title}{Entropy transport between two
 superconductors by electron tunneling}}.
\newblock {\emph{\JournalTitle{Phys. Rev. Lett.}}}
 \textbf{\bibinfo{volume}{15}}, \bibinfo{pages}{921--923}
 (\bibinfo{year}{1965}).

\bibitem{Gia06}
\bibinfo{author}{Giazotto, F.}, \bibinfo{author}{Heikkil\"a, T.~T.},
 \bibinfo{author}{Luukanen, A.}, \bibinfo{author}{Savin, A.~M.} \&
 \bibinfo{author}{Pekola, J.~P.}
\newblock \bibinfo{journal}{\bibinfo{title}{Opportunities for mesoscopics in
 thermometry and refrigeration: Physics and applications}}.
\newblock {\emph{\JournalTitle{Rev. Mod. Phys.}}}
 \textbf{\bibinfo{volume}{78}}, \bibinfo{pages}{217--274}
 (\bibinfo{year}{2006}).

\bibitem{MarSol14}
\bibinfo{author}{Mart{\'i}nez-P{\'e}rez, M.~J.}, \bibinfo{author}{Solinas, P.}
 \& \bibinfo{author}{Giazotto, F.}
\newblock \bibinfo{journal}{\bibinfo{title}{Coherent caloritronics in
 Josephson-based nanocircuits}}.
\newblock {\emph{\JournalTitle{J. Low Temp. Phys.}}}
 \textbf{\bibinfo{volume}{175}}, \bibinfo{pages}{813--837}
 (\bibinfo{year}{2014}).

\bibitem{ForGia17}
\bibinfo{author}{Fornieri, A.} \& \bibinfo{author}{Giazotto, F.}
\newblock \bibinfo{journal}{\bibinfo{title}{Towards phase-coherent
 caloritronics in superconducting circuits}}.
\newblock {\emph{\JournalTitle{Nat. Nanotechnology}}}
 \textbf{\bibinfo{volume}{12}}, \bibinfo{pages}{944--952}
 (\bibinfo{year}{2017}).

\bibitem{Gia12}
\bibinfo{author}{Giazotto, F.} \& \bibinfo{author}{Mart{\'\i}nez-P{\'e}rez,
 M.~J.}
\newblock \bibinfo{journal}{\bibinfo{title}{The Josephson heat
 interferometer}}.
\newblock {\emph{\JournalTitle{Nature}}} \textbf{\bibinfo{volume}{492}},
 \bibinfo{pages}{401--405} (\bibinfo{year}{2012}).

\bibitem{Mar14}
\bibinfo{author}{Mart{\'\i}nez-P{\'e}rez, M.~J.} \& \bibinfo{author}{Giazotto,
 F.}
\newblock \bibinfo{journal}{\bibinfo{title}{A quantum diffractor for thermal
 flux}}.
\newblock {\emph{\JournalTitle{Nat. Commun.}}} \textbf{\bibinfo{volume}{5}},
 \bibinfo{pages}{3579} (\bibinfo{year}{2014}).

\bibitem{Mar15}
\bibinfo{author}{Mart{\'\i}nez-P{\'e}rez, M.~J.}, \bibinfo{author}{Fornieri,
 A.} \& \bibinfo{author}{Giazotto, F.}
\newblock \bibinfo{journal}{\bibinfo{title}{Rectification of electronic heat
 current by a hybrid thermal diode}}.
\newblock {\emph{\JournalTitle{Nature Nanotechnology}}}
 \textbf{\bibinfo{volume}{10}}, \bibinfo{pages}{303--307}
 (\bibinfo{year}{2015}).

\bibitem{Sol16}
\bibinfo{author}{Solinas, P.}, \bibinfo{author}{Bosisio, R.} \&
 \bibinfo{author}{Giazotto, F.}
\newblock \bibinfo{journal}{\bibinfo{title}{Microwave quantum refrigeration
 based on the Josephson effect}}.
\newblock {\emph{\JournalTitle{Phys. Rev. B}}} \textbf{\bibinfo{volume}{93}},
 \bibinfo{pages}{224521} (\bibinfo{year}{2016}).

\bibitem{ForBla16}
\bibinfo{author}{Fornieri, A.}, \bibinfo{author}{Blanc, C.},
 \bibinfo{author}{Bosisio, R.}, \bibinfo{author}{D'Ambrosio, S.} \&
 \bibinfo{author}{Giazotto, F.}
\newblock \bibinfo{journal}{\bibinfo{title}{Nanoscale phase engineering of
 thermal transport with a Josephson heat modulator}}.
\newblock {\emph{\JournalTitle{Nat. Nanotechnology}}}
 \textbf{\bibinfo{volume}{11}}, \bibinfo{pages}{258--262}
 (\bibinfo{year}{2016}).

\bibitem{Pao17}
\bibinfo{author}{Paolucci, F.}, \bibinfo{author}{Marchegiani, G.},
 \bibinfo{author}{Strambini, E.} \& \bibinfo{author}{Giazotto, F.}
\newblock \bibinfo{journal}{\bibinfo{title}{Phase-tunable temperature
 amplifier}}.
\newblock {\emph{\JournalTitle{EPL (Europhysics Letters)}}}
 \textbf{\bibinfo{volume}{118}}, \bibinfo{pages}{68004}
 (\bibinfo{year}{2017}).

\bibitem{For17}
\bibinfo{author}{Fornieri, A.}, \bibinfo{author}{Timossi, G.},
 \bibinfo{author}{Virtanen, P.}, \bibinfo{author}{Solinas, P.} \&
 \bibinfo{author}{Giazotto, F.}
\newblock \bibinfo{journal}{\bibinfo{title}{0--$\pi$ phase-controllable thermal
 Josephson junction}}.
\newblock {\emph{\JournalTitle{Nat. Nanotechnology}}}
 \textbf{\bibinfo{volume}{12}}, \bibinfo{pages}{425--429}
 (\bibinfo{year}{2017}).

\bibitem{Pao18}
\bibinfo{author}{Paolucci, F.}, \bibinfo{author}{Marchegiani, G.},
 \bibinfo{author}{Strambini, E.} \& \bibinfo{author}{Giazotto, F.}
\newblock \bibinfo{journal}{\bibinfo{title}{Phase-tunable thermal logic:
 computation with heat}}.
\newblock {\emph{\JournalTitle{arXiv preprint arXiv:1709.08609}}}
 (\bibinfo{year}{2017}).

\bibitem{Tim18}
\bibinfo{author}{Timossi, G.~F.}, \bibinfo{author}{Fornieri, A.},
 \bibinfo{author}{Paolucci, F.}, \bibinfo{author}{Puglia, C.} \&
 \bibinfo{author}{Giazotto, F.}
\newblock \bibinfo{journal}{\bibinfo{title}{Phase-tunable Josephson thermal
 router}}.
\newblock {\emph{\JournalTitle{Nano Letters}}} \textbf{\bibinfo{volume}{18}},
 \bibinfo{pages}{1764--1769} (\bibinfo{year}{2018}).

\bibitem{GiaMar12}
\bibinfo{author}{Giazotto, F.} \& \bibinfo{author}{Mart{\'\i}nez-P{\'e}rez,
 M.~J.}
\newblock \bibinfo{journal}{\bibinfo{title}{Phase-controlled superconducting
 heat-flux quantum modulator}}.
\newblock {\emph{\JournalTitle{Appl. Phys. Lett.}}}
 \textbf{\bibinfo{volume}{101}}, \bibinfo{pages}{102601}
 (\bibinfo{year}{2012}).

\bibitem{Gia13}
\bibinfo{author}{Giazotto, F.}, \bibinfo{author}{Mart\'{\i}nez-P\'erez, M.~J.}
 \& \bibinfo{author}{Solinas, P.}
\newblock \bibinfo{journal}{\bibinfo{title}{Coherent diffraction of thermal
 currents in Josephson tunnel junctions}}.
\newblock {\emph{\JournalTitle{Phys. Rev. B}}} \textbf{\bibinfo{volume}{88}},
 \bibinfo{pages}{094506} (\bibinfo{year}{2013}).

\bibitem{Gua17}
\bibinfo{author}{Guarcello, C.}, \bibinfo{author}{Solinas, P.},
 \bibinfo{author}{Di~Ventra, M.} \& \bibinfo{author}{Giazotto, F.}
\newblock \bibinfo{journal}{\bibinfo{title}{Hysteretic superconducting
 heat-flux quantum modulator}}.
\newblock {\emph{\JournalTitle{Phys. Rev. Applied}}}
 \textbf{\bibinfo{volume}{7}}, \bibinfo{pages}{044021} (\bibinfo{year}{2017}).

\bibitem{GuaSol18}
\bibinfo{author}{Guarcello, C.}, \bibinfo{author}{Solinas, P.},
 \bibinfo{author}{Braggio, A.}, \bibinfo{author}{Di~Ventra, M.} \&
 \bibinfo{author}{Giazotto, F.}
\newblock \bibinfo{journal}{\bibinfo{title}{Josephson thermal memory}}.
\newblock {\emph{\JournalTitle{Phys. Rev. Applied}}}
 \textbf{\bibinfo{volume}{9}}, \bibinfo{pages}{014021} (\bibinfo{year}{2018}).

\bibitem{Gul07}
\bibinfo{author}{Gulevich, D.~R.} \& \bibinfo{author}{Kusmartsev, F.}
\newblock \bibinfo{journal}{\bibinfo{title}{New phenomena in long Josephson
 junctions}}.
\newblock {\emph{\JournalTitle{Supercond. Sci. Technol.}}}
 \textbf{\bibinfo{volume}{20}}, \bibinfo{pages}{S60} (\bibinfo{year}{2007}).

\bibitem{Mon12}
\bibinfo{author}{Monaco, R.}
\newblock \bibinfo{journal}{\bibinfo{title}{Magnetic sensors based on long
 Josephson tunnel junctions}}.
\newblock {\emph{\JournalTitle{Supercond. Sci. Technol.}}}
 \textbf{\bibinfo{volume}{25}}, \bibinfo{pages}{115011}
 (\bibinfo{year}{2012}).

\bibitem{Val14}
\bibinfo{author}{Valenti, D.}, \bibinfo{author}{Guarcello, C.} \&
 \bibinfo{author}{Spagnolo, B.}
\newblock \bibinfo{journal}{\bibinfo{title}{Switching times in long-overlap
 Josephson junctions subject to thermal fluctuations and non-Gaussian noise
 sources}}.
\newblock {\emph{\JournalTitle{Phys. Rev. B}}} \textbf{\bibinfo{volume}{89}},
 \bibinfo{pages}{214510} (\bibinfo{year}{2014}).

\bibitem{Gua15}
\bibinfo{author}{Guarcello, C.}, \bibinfo{author}{Valenti, D.},
 \bibinfo{author}{Carollo, A.} \& \bibinfo{author}{Spagnolo, B.}
\newblock \bibinfo{journal}{\bibinfo{title}{Stabilization effects of
 dichotomous noise on the lifetime of the superconducting state in a long
 Josephson junction}}.
\newblock {\emph{\JournalTitle{Entropy}}} \textbf{\bibinfo{volume}{17}},
 \bibinfo{pages}{2862} (\bibinfo{year}{2015}).


\bibitem{Zel15}
\bibinfo{author}{Zelikman, M.~A.}
\newblock \bibinfo{journal}{\bibinfo{title}{Hysteresis in the behavior of a
 long modulated Josephson junction in a magnetic field for small values of the
 pinning parameter}}.
\newblock {\emph{\JournalTitle{Technical Physics}}}
 \textbf{\bibinfo{volume}{60}}, \bibinfo{pages}{1299--1304}
 (\bibinfo{year}{2015}).

\bibitem{Pan15}
\bibinfo{author}{Pankratov, A.~L.}, \bibinfo{author}{Fedorov, K.~G.},
 \bibinfo{author}{Salerno, M.}, \bibinfo{author}{Shitov, S.~V.} \&
 \bibinfo{author}{Ustinov, A.~V.}
\newblock \bibinfo{journal}{\bibinfo{title}{Nonreciprocal transmission of
 microwaves through a long Josephson junction}}.
\newblock {\emph{\JournalTitle{Phys. Rev. B}}} \textbf{\bibinfo{volume}{92}},
 \bibinfo{pages}{104501} (\bibinfo{year}{2015}).

\bibitem{GuaValSpa16}
\bibinfo{author}{Guarcello, C.}, \bibinfo{author}{Valenti, D.},
 \bibinfo{author}{Carollo, A.} \& \bibinfo{author}{Spagnolo, B.}
\newblock \bibinfo{journal}{\bibinfo{title}{Effects of L\'evy noise on the
 dynamics of sine-Gordon solitons in long Josephson junctions}}.
\newblock {\emph{\JournalTitle{J. Stat. Mech.: Theory Exp.}}}
 \textbf{\bibinfo{volume}{2016}}, \bibinfo{pages}{054012}
 (\bibinfo{year}{2016}).

\bibitem{GuaSol17}
\bibinfo{author}{Guarcello, C.}, \bibinfo{author}{Solinas, P.},
 \bibinfo{author}{Di~Ventra, M.} \& \bibinfo{author}{Giazotto, F.}
\newblock \bibinfo{journal}{\bibinfo{title}{Solitonic Josephson-based
 meminductive systems}}.
\newblock {\emph{\JournalTitle{Sci. Rep.}}} \textbf{\bibinfo{volume}{7}},
 \bibinfo{pages}{46736} (\bibinfo{year}{2017}).

\bibitem{Hil18}
\bibinfo{author}{Hill, D.}, \bibinfo{author}{Kim, S.~K.} \&
 \bibinfo{author}{Tserkovnyak, Y.}
\newblock \bibinfo{journal}{\bibinfo{title}{Spin--Torque--Biased Magnetic Strip: Nonequilibrium Phase Diagram and Relation to Long Josephson Junctions}}.
\newblock {\emph{\JournalTitle{Phys. Rev. Lett.}}}
 \textbf{\bibinfo{volume}{121}}, \bibinfo{pages}{037202}
 (\bibinfo{year}{2018}).

\bibitem{Ooi07}
\bibinfo{author}{Ooi, S.} \emph{et~al.}
\newblock \bibinfo{journal}{\bibinfo{title}{Nonlinear nanodevices using
 magnetic flux quanta}}.
\newblock {\emph{\JournalTitle{Phys. Rev. Lett.}}}
 \textbf{\bibinfo{volume}{99}}, \bibinfo{pages}{207003}
 (\bibinfo{year}{2007}).

\bibitem{Fed12}
\bibinfo{author}{Fedorov, K.~G.}, \bibinfo{author}{Shitov, S.~V.},
 \bibinfo{author}{Rotzinger, H.} \& \bibinfo{author}{Ustinov, A.~V.}
\newblock \bibinfo{journal}{\bibinfo{title}{Nonreciprocal microwave
 transmission through a long Josephson junction}}.
\newblock {\emph{\JournalTitle{Phys. Rev. B}}} \textbf{\bibinfo{volume}{85}},
 \bibinfo{pages}{184512} (\bibinfo{year}{2012}).

\bibitem{Mon13}
\bibinfo{author}{Monaco, R.}, \bibinfo{author}{Granata, C.},
 \bibinfo{author}{Russo, R.} \& \bibinfo{author}{Vettoliere, A.}
\newblock \bibinfo{journal}{\bibinfo{title}{Ultra-low-noise magnetic sensing
 with long Josephson tunnel junctions}}.
\newblock {\emph{\JournalTitle{Supercond. Sci. Technol.}}}
 \textbf{\bibinfo{volume}{26}}, \bibinfo{pages}{125005}
 (\bibinfo{year}{2013}).

\bibitem{Gra14}
\bibinfo{author}{Granata, C.}, \bibinfo{author}{Vettoliere, A.} \&
 \bibinfo{author}{Monaco, R.}
\newblock \bibinfo{journal}{\bibinfo{title}{Noise performance of
 superconductive magnetometers based on long Josephson tunnel junctions}}.
\newblock {\emph{\JournalTitle{Supercond. Sci. Technol.}}}
 \textbf{\bibinfo{volume}{27}}, \bibinfo{pages}{095003}
 (\bibinfo{year}{2014}).

\bibitem{Kos14}
\bibinfo{author}{Koshelets, V.~P.}
\newblock \bibinfo{journal}{\bibinfo{title}{Sub-terahertz sound excitation and
 detection by a long Josephson junction}}.
\newblock {\emph{\JournalTitle{Supercond. Sci. Technol.}}}
 \textbf{\bibinfo{volume}{27}}, \bibinfo{pages}{065010}
 (\bibinfo{year}{2014}).

\bibitem{Fed14}
\bibinfo{author}{Fedorov, K.~G.}, \bibinfo{author}{Shcherbakova, A.~V.},
 \bibinfo{author}{Wolf, M.~J.}, \bibinfo{author}{Beckmann, D.} \&
 \bibinfo{author}{Ustinov, A.~V.}
\newblock \bibinfo{journal}{\bibinfo{title}{Fluxon readout of a superconducting
 qubit}}.
\newblock {\emph{\JournalTitle{Phys. Rev. Lett.}}}
 \textbf{\bibinfo{volume}{112}}, \bibinfo{pages}{160502}
 (\bibinfo{year}{2014}).

\bibitem{Vet15}
\bibinfo{author}{Vettoliere, A.}, \bibinfo{author}{Granata, C.} \&
 \bibinfo{author}{Monaco, R.}
\newblock \bibinfo{journal}{\bibinfo{title}{Long Josephson junction in
 ultralow-noise magnetometer configuration}}.
\newblock {\emph{\JournalTitle{IEEE Trans. Magn.}}}
 \textbf{\bibinfo{volume}{51}}, \bibinfo{pages}{1--4} (\bibinfo{year}{2015}).

\bibitem{Gol17}
\bibinfo{author}{Golovchanskiy, I.~A.} \emph{et~al.}
\newblock \bibinfo{journal}{\bibinfo{title}{Ferromagnetic resonance with long
 Josephson junction}}.
\newblock {\emph{\JournalTitle{Supercond. Sci. Technol.}}}
 \textbf{\bibinfo{volume}{30}}, \bibinfo{pages}{054005}
 (\bibinfo{year}{2017}).

\bibitem{Bar82}
\bibinfo{author}{Barone, A.} \& \bibinfo{author}{Patern\`{o}, G.}
\newblock \emph{\bibinfo{title}{Physics and Applications of the Josephson
 Effect}} (\bibinfo{publisher}{Wiley, New York}, \bibinfo{year}{1982}).

\bibitem{Tin04}
\bibinfo{author}{Tinkham, M.}
\newblock \emph{\bibinfo{title}{Introduction to Superconductivity: Second
 Edition}}.
\newblock Dover Books on Physics (\bibinfo{publisher}{Dover Publications},
 \bibinfo{year}{2004}).

\bibitem{Par93}
\bibinfo{author}{Parmentier, R.~D.}
\newblock \emph{\bibinfo{title}{Solitons and Long Josephson Junctions}},
 \bibinfo{pages}{221--248} (\bibinfo{publisher}{Springer Netherlands},
 \bibinfo{address}{Dordrecht}, \bibinfo{year}{1993}).

\bibitem{Ust98}
\bibinfo{author}{Ustinov, A.~V.}
\newblock \bibinfo{journal}{\bibinfo{title}{Solitons in Josephson junctions}}.
\newblock {\emph{\JournalTitle{Physica D}}} \textbf{\bibinfo{volume}{123}},
 \bibinfo{pages}{315--329} (\bibinfo{year}{1998}).

\bibitem{Gold01}
\bibinfo{author}{Goldobin, E.}
\newblock \bibinfo{title}{Flux-flow oscillators and phenomenon of Cherenkov
 radiation from fast moving fluxons}.
\newblock In \emph{\bibinfo{booktitle}{Microwave Superconductivity}},
 \bibinfo{pages}{581--614} (\bibinfo{publisher}{Springer},
 \bibinfo{year}{2001}).
 
 \bibitem{Wel94}
\bibinfo{author}{Wellstood, F.~C.}, \bibinfo{author}{Urbina, C.} \&
 \bibinfo{author}{Clarke, J.}
\newblock \bibinfo{journal}{\bibinfo{title}{Hot-electron effects in metals}}.
\newblock {\emph{\JournalTitle{Phys. Rev. B}}} \textbf{\bibinfo{volume}{49}},
 \bibinfo{pages}{5942--5955} (\bibinfo{year}{1994}).



\bibitem{Gia05}
\bibinfo{author}{Giazotto, F.} \& \bibinfo{author}{Pekola, J.~P.}
\newblock \bibinfo{journal}{\bibinfo{title}{Josephson tunnel junction
 controlled by quasiparticle injection}}.
\newblock {\emph{\JournalTitle{J. Appl. Phys.}}} \textbf{\bibinfo{volume}{97}}
 (\bibinfo{year}{2005}).

\bibitem{Bos16}
\bibinfo{author}{Bosisio, R.}, \bibinfo{author}{Solinas, P.},
 \bibinfo{author}{Braggio, A.} \& \bibinfo{author}{Giazotto, F.}
\newblock \bibinfo{journal}{\bibinfo{title}{Photonic heat conduction in
 Josephson-coupled Bardeen-Cooper-Schrieffer superconductors}}.
\newblock {\emph{\JournalTitle{Phys. Rev. B}}} \textbf{\bibinfo{volume}{93}},
 \bibinfo{pages}{144512} (\bibinfo{year}{2016}).

\bibitem{Gua16}
\bibinfo{author}{Guarcello, C.}, \bibinfo{author}{Giazotto, F.} \&
 \bibinfo{author}{Solinas, P.}
\newblock \bibinfo{journal}{\bibinfo{title}{Coherent diffraction of thermal
 currents in long Josephson tunnel junctions}}.
\newblock {\emph{\JournalTitle{Phys. Rev. B}}} \textbf{\bibinfo{volume}{94}},
 \bibinfo{pages}{054522} (\bibinfo{year}{2016}).

\bibitem{Cir97}
\bibinfo{author}{Cirillo, M.}, \bibinfo{author}{Doderer, T.},
 \bibinfo{author}{Lachenmann, S.~G.}, \bibinfo{author}{Santucci, F.} \&
 \bibinfo{author}{Gr\o{}nbech-Jensen, N.}
\newblock \bibinfo{journal}{\bibinfo{title}{Dynamical evidence of critical
 fields in josephson junctions}}.
\newblock {\emph{\JournalTitle{Phys. Rev. B}}} \textbf{\bibinfo{volume}{56}},
 \bibinfo{pages}{11889--11896} (\bibinfo{year}{1997}).


\bibitem{Kup06}
\bibinfo{author}{Kuplevakhsky, S.~V.} \& \bibinfo{author}{Glukhov, A.~M.}
\newblock \bibinfo{journal}{\bibinfo{title}{Static solitons of the sine-Gordon
 equation and equilibrium vortex structure in Josephson junctions}}.
\newblock {\emph{\JournalTitle{Phys. Rev. B}}} \textbf{\bibinfo{volume}{73}},
 \bibinfo{pages}{024513} (\bibinfo{year}{2006}).

\bibitem{Kup07}
\bibinfo{author}{Kuplevakhsky, S.~V.} \& \bibinfo{author}{Glukhov, A.~M.}
\newblock \bibinfo{journal}{\bibinfo{title}{Exact analytical solution of the
 problem of current-carrying states of the josephson junction in external
 magnetic fields}}.
\newblock {\emph{\JournalTitle{Phys. Rev. B}}} \textbf{\bibinfo{volume}{76}},
 \bibinfo{pages}{174515} (\bibinfo{year}{2007}).


\bibitem{Kup10}
\bibinfo{author}{Kuplevakhsky, S.~V.} \& \bibinfo{author}{Glukhov, A.~M.}
\newblock \bibinfo{journal}{\bibinfo{title}{Exact analytical solution of a
 classical josephson tunnel junction problem}}.
\newblock {\emph{\JournalTitle{Low Temp. Phys.}}}
 \textbf{\bibinfo{volume}{36}}, \bibinfo{pages}{1012--1021}
 (\bibinfo{year}{2010}).

\bibitem{Che94}
\bibinfo{author}{Chen, D.-X.} \& \bibinfo{author}{Hernando, A.}
\newblock \bibinfo{journal}{\bibinfo{title}{Magnetization of uniform josephson
 junctions}}.
\newblock {\emph{\JournalTitle{Phys. Rev. B}}} \textbf{\bibinfo{volume}{49}},
 \bibinfo{pages}{465--474} (\bibinfo{year}{1994}).

\bibitem{Yug95}
\bibinfo{author}{Yugay, K.~N.}, \bibinfo{author}{Blinov, N.~V.} \&
 \bibinfo{author}{Shirokov, I.~V.}
\newblock \bibinfo{journal}{\bibinfo{title}{Effect of memory and dynamical
 chaos in long josephson junctions}}.
\newblock {\emph{\JournalTitle{Phys. Rev. B}}} \textbf{\bibinfo{volume}{51}},
 \bibinfo{pages}{12737--12741} (\bibinfo{year}{1995}).

\bibitem{Yug99}
\bibinfo{author}{Yugay, K.~N.}, \bibinfo{author}{Blinov, N.~V.} \&
 \bibinfo{author}{Shirokov, I.~V.}
\newblock \bibinfo{journal}{\bibinfo{title}{Bifurcations and a chaos strip in
 states of long josephson junctions}}.
\newblock {\emph{\JournalTitle{Low Temp. Phys.}}}
 \textbf{\bibinfo{volume}{25}}, \bibinfo{pages}{530--534}
 (\bibinfo{year}{1999}).


\bibitem{Hue88}
\bibinfo{author}{Huebener, R.}
\newblock \bibinfo{title}{Scanning electron microscopy at very low
 temperatures}.
\newblock vol.~\bibinfo{volume}{70} of \emph{\bibinfo{series}{Advances in
 Electronics and Electron Physics}}, \bibinfo{pages}{1 -- 78}
 (\bibinfo{publisher}{Academic Press}, \bibinfo{year}{1988}).

\bibitem{Mal94}
\bibinfo{author}{Malomed, B.~A.} \& \bibinfo{author}{Ustinov, A.~V.}
\newblock \bibinfo{journal}{\bibinfo{title}{Analysis of testing the
 single-fluxon dynamics in a long Josephson junction by a dissipative spot}}.
\newblock {\emph{\JournalTitle{Phys. Rev. B}}} \textbf{\bibinfo{volume}{49}},
 \bibinfo{pages}{13024--13029} (\bibinfo{year}{1994}).

\bibitem{Dod97}
\bibinfo{author}{Doderer, T.}
\newblock \bibinfo{journal}{\bibinfo{title}{Microscopic imaging of Josephson
 junction dynamics}}.
\newblock {\emph{\JournalTitle{Int. J. Mod. Phys. B}}}
 \textbf{\bibinfo{volume}{11}}, \bibinfo{pages}{1979--2042}
 (\bibinfo{year}{1997}).

\bibitem{Gas15}
\bibinfo{author}{Gasparinetti, S.} \emph{et~al.}
\newblock \bibinfo{journal}{\bibinfo{title}{Fast electron thermometry for
 ultrasensitive calorimetric detection}}.
\newblock {\emph{\JournalTitle{Phys. Rev. Applied}}}
 \textbf{\bibinfo{volume}{3}}, \bibinfo{pages}{014007} (\bibinfo{year}{2015}).

\bibitem{Gia15}
\bibinfo{author}{Giazotto, F.}, \bibinfo{author}{Solinas, P.},
 \bibinfo{author}{Braggio, A.} \& \bibinfo{author}{Bergeret, F.~S.}
\newblock \bibinfo{journal}{\bibinfo{title}{Ferromagnetic-insulator-based
 superconducting junctions as sensitive electron thermometers}}.
\newblock {\emph{\JournalTitle{Phys. Rev. Applied}}}
 \textbf{\bibinfo{volume}{4}}, \bibinfo{pages}{044016} (\bibinfo{year}{2015}).

\bibitem{Sai16}
\bibinfo{author}{Saira, O.-P.}, \bibinfo{author}{Zgirski, M.},
 \bibinfo{author}{Viisanen, K.~L.}, \bibinfo{author}{Golubev, D.~S.} \&
 \bibinfo{author}{Pekola, J.~P.}
\newblock \bibinfo{journal}{\bibinfo{title}{Dispersive thermometry with a
 Josephson junction coupled to a resonator}}.
\newblock {\emph{\JournalTitle{Phys. Rev. Applied}}}
 \textbf{\bibinfo{volume}{6}}, \bibinfo{pages}{024005} (\bibinfo{year}{2016}).

\bibitem{Zgi17}
\bibinfo{author}{Zgirski, M.}, \bibinfo{author}{Foltyn, M.},
 \bibinfo{author}{Savin, A.}, \bibinfo{author}{Meschke, M.} \&
 \bibinfo{author}{Pekola, J.}
\newblock \bibinfo{journal}{\bibinfo{title}{Nanosecond thermometry with
 Josephson junction}}.
\newblock {\emph{\JournalTitle{arXiv preprint arXiv:1704.04762}}}
 (\bibinfo{year}{2017}).

\bibitem{Wan18}
\bibinfo{author}{Wang, L.~B.}, \bibinfo{author}{Saira, O.-P.} \&
 \bibinfo{author}{Pekola, J.~P.}
\newblock \bibinfo{journal}{\bibinfo{title}{Fast thermometry with a proximity
 Josephson junction}}.
\newblock {\emph{\JournalTitle{Appl. Phys. Lett.}}}
 \textbf{\bibinfo{volume}{112}}, \bibinfo{pages}{013105}
 (\bibinfo{year}{2018}).

\bibitem{Vir18}
\bibinfo{author}{Virtanen, P.}, \bibinfo{author}{Ronzani, A.} \&
 \bibinfo{author}{Giazotto, F.}
\newblock \bibinfo{journal}{\bibinfo{title}{Josephson photodetectors via
 temperature-to-phase conversion}}.
\newblock {\emph{\JournalTitle{Phys. Rev. Applied}}}
 \textbf{\bibinfo{volume}{9}}, \bibinfo{pages}{054027} (\bibinfo{year}{2018}).


\bibitem{Kim11}
\bibinfo{author}{Kim, S.~W.}, \bibinfo{author}{Sagawa, T.},
 \bibinfo{author}{De~Liberato, S.} \& \bibinfo{author}{Ueda, M.}
\newblock \bibinfo{journal}{\bibinfo{title}{Quantum Szilard engine}}.
\newblock {\emph{\JournalTitle{Phys. Rev. Lett.}}}
 \textbf{\bibinfo{volume}{106}}, \bibinfo{pages}{070401}
 (\bibinfo{year}{2011}).

\bibitem{Ros14}
\bibinfo{author}{Ro\ss{}nagel, J.}, \bibinfo{author}{Abah, O.},
 \bibinfo{author}{Schmidt-Kaler, F.}, \bibinfo{author}{Singer, K.} \&
 \bibinfo{author}{Lutz, E.}
\newblock \bibinfo{journal}{\bibinfo{title}{Nanoscale heat engine beyond the
 Carnot limit}}.
\newblock {\emph{\JournalTitle{Phys. Rev. Lett.}}}
 \textbf{\bibinfo{volume}{112}}, \bibinfo{pages}{030602}
 (\bibinfo{year}{2014}).

\bibitem{Ros16}
\bibinfo{author}{Ro{\ss}nagel, J.} \emph{et~al.}
\newblock \bibinfo{journal}{\bibinfo{title}{A single-atom heat engine}}.
\newblock {\emph{\JournalTitle{Science}}} \textbf{\bibinfo{volume}{352}},
 \bibinfo{pages}{325--329} (\bibinfo{year}{2016}).

\bibitem{Cam16}
\bibinfo{author}{Campisi, M.} \& \bibinfo{author}{Fazio, R.}
\newblock \bibinfo{journal}{\bibinfo{title}{The power of a critical heat
 engine}}.
\newblock {\emph{\JournalTitle{Nat. Commun.}}} \textbf{\bibinfo{volume}{7}},
 \bibinfo{pages}{11895} (\bibinfo{year}{2016}).

\bibitem{Mar16}
\bibinfo{author}{Marchegiani, G.}, \bibinfo{author}{Virtanen, P.},
 \bibinfo{author}{Giazotto, F.} \& \bibinfo{author}{Campisi, M.}
\newblock \bibinfo{journal}{\bibinfo{title}{Self-oscillating Josephson quantum
 heat engine}}.
\newblock {\emph{\JournalTitle{Phys. Rev. Applied}}}
 \textbf{\bibinfo{volume}{6}}, \bibinfo{pages}{054014} (\bibinfo{year}{2016}).

\bibitem{Gol13}
\bibinfo{author}{Golubev, D.}, \bibinfo{author}{Faivre, T.} \&
 \bibinfo{author}{Pekola, J.~P.}
\newblock \bibinfo{journal}{\bibinfo{title}{Heat transport through a Josephson
 junction}}.
\newblock {\emph{\JournalTitle{Phys. Rev. B}}} \textbf{\bibinfo{volume}{87}},
 \bibinfo{pages}{094522} (\bibinfo{year}{2013}).

\bibitem{Dyn78}
\bibinfo{author}{Dynes, R.~C.}, \bibinfo{author}{Narayanamurti, V.} \&
 \bibinfo{author}{Garno, J.~P.}
\newblock \bibinfo{journal}{\bibinfo{title}{Direct measurement of
 quasiparticle-lifetime broadening in a strong-coupled superconductor}}.
\newblock {\emph{\JournalTitle{Phys. Rev. Lett.}}}
 \textbf{\bibinfo{volume}{41}}, \bibinfo{pages}{1509--1512}
 (\bibinfo{year}{1978}).

\bibitem{Vir17}
\bibinfo{author}{Virtanen, P.}, \bibinfo{author}{Solinas, P.} \&
 \bibinfo{author}{Giazotto, F.}
\newblock \bibinfo{journal}{\bibinfo{title}{Spectral representation of the heat
 current in a driven josephson junction}}.
\newblock {\emph{\JournalTitle{Phys. Rev. B}}} \textbf{\bibinfo{volume}{95}},
 \bibinfo{pages}{144512} (\bibinfo{year}{2017}).

\bibitem{Pek09}
\bibinfo{author}{Timofeev, A.~V.} \emph{et~al.}
\newblock \bibinfo{journal}{\bibinfo{title}{Recombination-limited energy
 relaxation in a Bardeen-Cooper-Schrieffer superconductor}}.
\newblock {\emph{\JournalTitle{Phys. Rev. Lett.}}}
 \textbf{\bibinfo{volume}{102}}, \bibinfo{pages}{017003}
 (\bibinfo{year}{2009}).

\end{thebibliography}

\section*{Acknowledgements}

We acknowledge P. Virtanen for fruitful discussions.
C.G., A.B., and F.G. acknowledge the European Research Council under the European Union's Seventh Framework Program (FP7/2007-2013)/ERC Grant agreement No.~615187-COMANCHE and the Tuscany Region under the FARFAS 2014 project SCIADRO for partial financial support. 
P.S. and A.B. have received funding from the European Union FP7/2007-2013 under REA Grant agreement No. 630925 -- COHEAT. 
A.B. acknowledges the CNR-CONICET cooperation programme ``Energy conversion in quantum nanoscale hybrid devices'' and the Royal Society though the International Exchanges between the UK and Italy (grant IES R3 170054).

\section*{Author contributions statement}

C.G., P.S., A.B., and F.G. conceived the ideas and designed the study. C.G. developed the algorithm and carried out the simulations. C.G., P.S., A.B., and F.G discussed the results and wrote the paper.

\section*{Additional information}

The authors declare no competing interests.

\end{document}